\documentclass[lettersize,journal]{IEEEtran}
\usepackage{amsmath,amsfonts}
\usepackage{algorithmic}
\usepackage{array}
\usepackage[caption=false,font=normalsize,labelfont=sf,textfont=sf]{subfig}
\usepackage{textcomp}
\usepackage{booktabs}
\usepackage{tabularx, booktabs}
\usepackage{stfloats}
\usepackage{url}
\usepackage{verbatim}
\usepackage{graphicx}
\usepackage{tikz}
\usepackage{xcolor} 
\usetikzlibrary{arrows.meta, positioning}
\def\BibTeX{{\rm B\kern-.05em{\sc i\kern-.025em b}\kern-.08em T\kern-.1667em\lower.7ex\hbox{E}\kern-.125emX}}

\begin{document}

\title{Reinforcement Learning based Traffic Signal Control for Queue Minimization: From Lane-based to Laneless Traffic}
\author{Anirud Nandakumar$^{1,2}$, Chayan Banerjee$^{3,*}$, Lelitha Devi Vanajakshi$^{1,2}$
\thanks{$^{1}$ Department of Civil Engineering, IIT Madras}%
\thanks{$^{2}$ Wadhwani School of Data Science \& AI, IIT Madras}%
\thanks{$^{3}$ School of Electrical Engineering \& Robotics, QUT, Brisbane, Australia}%
\thanks{\hspace{0pt}\parbox{\textwidth}{Corresponding author: Chayan Banerjee (c.banerjee@qut.edu.au)}}%
}
\maketitle

\begin{abstract}
Efficient traffic signal control (TSC) is essential for reducing congestion and delays. Traditional TSC methods struggle to capture the dynamics of laneless, heterogeneous traffic. We propose a reinforcement learning (RL) based adaptive TSC framework whose reward depends only on approach-level queue length, motivated by the difficulty of obtaining per-vehicle measurements in laneless traffic, and evaluate it using SUMO on a conventional lane-based, homogeneous (LBHO) intersection and on a non-orthogonal laneless, heterogeneous (LLHE) intersection. On the LBHO site, the lowest mean configuration has about 23\% lower mean queue length than a Dynamic Webster baseline, and the unselected Full-State configuration about 15\% lower. On the LLHE site, all RL configurations have 66--73\% lower mean queue length than a Dynamic Webster baseline under nominal demand, and 63--74\% lower under two higher-demand scenarios evaluated without retraining. Our algorithm also outperforms actuated control and a RESCO-DQN baseline on mean queue length. We additionally release a laneless, heterogeneous-vehicle SUMO framework integrated with RL based signal control, to our knowledge the first such open implementation on a current SUMO version.
\end{abstract}


\section{Introduction}
Traffic signal control (TSC) is essential for managing traffic flows, ensuring road safety, and reducing delays, pollution, and economic losses. Traditional approaches, including fixed-time, actuated, and adaptive control methods, each offer progressively greater responsiveness to real-time conditions, yet remain limited in their ability to handle highly dynamic or non-standard traffic patterns.

These limitations have motivated the use of reinforcement learning (RL) for adaptive TSC, as RL can learn effective control strategies directly from interaction with the environment without relying on predefined traffic models. However, most existing RL based approaches optimize objectives such as delay or network pressure, which require vehicle specific measurements that are difficult to obtain under laneless, heterogeneous traffic conditions common in countries such as India.

\begin{figure}
    \centering
    \includegraphics[width=1\linewidth]{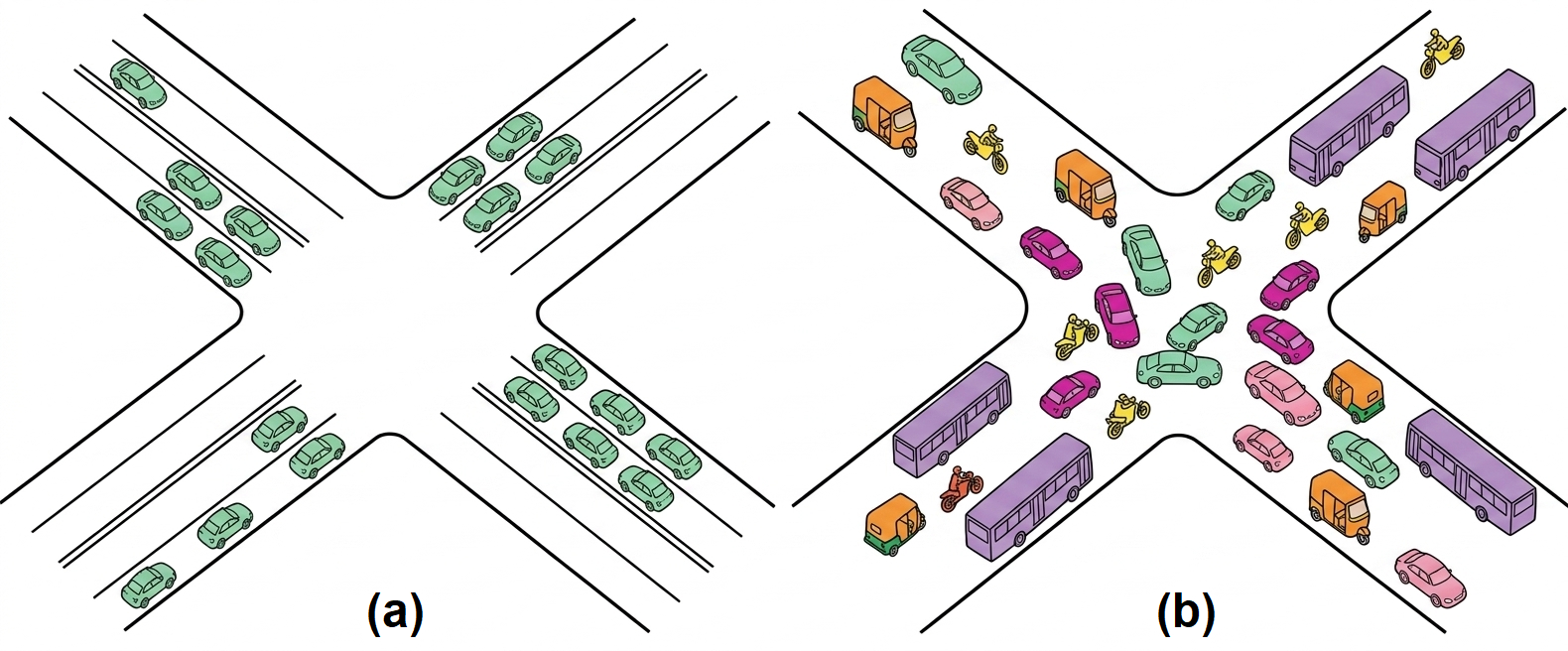}
    \caption{A comparative illustration of traffic behavior at a four-way intersection: (a) represents highly structured, lane based traffic with a homogeneous fleet of vehicles (LBHO) following road markings, while (b) depicts chaotic, laneless, heterogeneous traffic (LLHE) featuring a mixed variety of vehicles (buses, cars, rickshaws, and motorcycles) navigating the same space without order.}
    \label{fig:traffic_comparison}
\end{figure}

Under such conditions, vehicles of widely varying size and manoeuvrability share the same road width,  without strictly adhering to lane markings, making per-lane, per-vehicle observations unreliable (Fig.~\ref{fig:traffic_comparison}). This motivates control strategies built on more robust, aggregate traffic descriptors such as queue length, which can be estimated using cameras or data-driven models \cite{li2015review, seo2017traffic}.

This paper proposes an adaptive RL based TSC framework that directly minimizes total queue length across all signal phases. The framework is applied, with separately trained controllers, to a conventional lane-based intersection with a homogeneous vehicle fleet, hereafter the lane-based, homogeneous (LBHO) site, and to a real, non-orthogonal junction carrying laneless, heterogeneous traffic, hereafter the laneless, heterogeneous (LLHE) site. At the LLHE site, controllers trained on nominal demand are additionally evaluated, without retraining, under two higher-demand scenarios.

\noindent
The primary contributions of this paper are listed as follows:

\begin{itemize}
\item \textbf{Infrastructure:} an open-source LLHE SUMO framework integrated with RL based signal control; to our knowledge, the first such implementation on a current SUMO version.

\item \textbf{Queue based adaptive control:} an RL
traffic signal controller whose reward relies only on approach-level
queue length, applicable from conventional lane-based homogeneous (LBHO) to laneless,
heterogeneous traffic (LLHE).

\item \textbf{Cross-regime evaluation:} an empirical
study on an LBHO and an LLHE intersection, comparing the
controller with a classical (dynamic Webster) baseline, alternative reward formulations,
and several state representations.
\end{itemize}

The remainder of this paper is organized as follows. Section~\ref{sec:lit} reviews related work. Section~\ref{sec:prelim} presents the RL preliminaries, followed by the problem formulation in Section~\ref{sec:formulation}. Section~\ref{sec:method} details the proposed methodology, and Section~\ref{sec:laneless-sumo} describes the LLHE SUMO simulation framework. Sections~\ref{sec:setup} and~\ref{sec:baselines} describe the experimental setup and baselines, respectively. Section~\ref{sec:results} presents and discusses the results, and Section~\ref{sec:conclusion} concludes with directions for future work.

\noindent
The code associated with this paper is publicly available.\footnote{\url{https://github.com/AnirudN/RLTSCQ}}

\section{Literature Review} \label{sec:lit}
TSC strategies have evolved from empirical, fixed-time schemes to
data-driven adaptive controllers. Fixed-time methods such as
Webster's~\cite{webster1958traffic} compute cycle length and green
splits from historical flow and are easy to deploy, but cannot respond
to real-time variation. Actuated control~\cite{bullen1989effects}
extends or skips phases based on local detectors, improving
responsiveness without network-level
coordination~\cite{koonce2008traffic}. Adaptive systems such as
SCATS~\cite{sims1980sydney} and SCOOT~\cite{hunt1982scoot} coordinate
across intersections using traffic models, but their reliance on
per-lane flow assumptions limits their applicability under
LLHE conditions.

These limitations motivated the shift to learning based control.
Reinforcement learning (RL) offers model-free adaptation directly from
interaction with the environment~\cite{sutton1998reinforcement}. Early
tabular Q-learning studies established feasibility for TSC but did
not scale~\cite{abdulhai2003reinforcement}, motivating deep RL
approaches such as DQN~\cite{mnih2015dqn} and asynchronous
actor--critic methods~\cite{mnih2016a3c}. Among these, Proximal Policy
Optimization (PPO)~\cite{schulman2017proximal} has been widely adopted in recent RL-TSC work owing to its stable
clipped updates and sample efficiency under the noisy dynamics
typical of signalized intersections; it is adopted throughout the
present work and described further in Section~\ref{sec:ppo}. Recent
surveys~\cite{noaeen2022reinforcement, wei2021survey} and
standardized benchmarks such as
RESCO~\cite{NEURIPSDATASETSANDBENCHMARKS2021_f0935e4c} have
consolidated this line of work, and network-aware architectures such
as CoLight~\cite{wei2019colight} and MPLight~\cite{chen2020mplight}
have extended it to multi-intersection coordination.

\emph{Reward design:} RL-TSC rewards reported in the literature
typically target a single instantaneous congestion measure:
delay~\cite{liang2019deep},
max-pressure~\cite{varaiya2013mp, wei2019presslight}, or average
speed~\cite{egea2020assessment}. Queue based rewards have also been
used, most notably as one term in the composite reward of
IntelliLight~\cite{wei2018intellilight}, and as a primary reward signal in
subsequent heterogeneous-traffic
work~\cite{RM2023100124, du2022rltsc}. This paper differs in that
its queue based reward combines an absolute congestion term with a
queue reduction term (see Section~\ref{sec:formulation}); in 
Section~\ref{subsec:alt-rewards} we compare this formulation
empirically against the delay-, pressure-, and speed based
alternatives above.

\emph{State representation:} State-abstraction theory establishes
that an abstraction improves RL performance only insofar as it
preserves task-relevant information~\cite{li2006unified,
abel2016near}, and autoencoder based compression is a well-established instance of this trade-off~\cite{lange2010deep}. Near-universal RL-TSC practice already includes signal-phase as a state feature for essentially this reason. Autoencoder encodings are used as one of several state representations whose effect on queue length is reported empirically (Section~\ref{sec:results}).

\emph{Heterogeneous and laneless traffic:} Most RL-TSC work has
targeted lane based traffic with a homogeneous vehicle fleet. Recent
work has begun to address heterogeneous
conditions~\cite{RM2023100124, zhang2024heterolight, du2022rltsc},
but existing studies typically retain lane discipline, target
older simulator versions without an RL
integration~\cite{patel2016comparison}, or focus on connected/robot
vehicles in an otherwise lane based
setting~\cite{wang2024mixedtraffic}. RL-TSC for genuinely
\emph{laneless} heterogeneous traffic, where vehicles do
not adhere to lane markings, remains comparatively underexplored, and
we are not aware of an openly released implementation combining a
laneless, heterogeneous-vehicle configuration on a current SUMO
version with RL based signal control. Under such conditions,
macroscopic descriptors such as total queue length are more practical
than per-vehicle metrics, which motivates the reward design used
here. This paper addresses this gap by introducing a
queue-length based RL-TSC framework, first on a conventional
lane based intersection and then on a real heterogeneous, laneless
site, \textcolor{black}{the latter additionally evaluated under two higher-demand scenarios without retraining.}

\section{Preliminaries}\label{sec:prelim}
RL is a machine learning framework wherein a decision-maker (the controller) observes the current state of a dynamic environment and learns to take actions that maximize a cumulative reward. The ultimate goal is to learn an optimal policy that yields the highest expected reward over time. Mathematically, these problems are formulated as Markov Decision Processes (MDPs), which model a controller interacting with an environment over discrete time steps: at each step the controller chooses an action, the environment moves to a new state that depends only on the current state and that action, and the controller receives a scalar reward. In this framework, TSC operates as a continuing task, meaning the controller must continuously execute actions without a predefined endpoint.\newline

\noindent
MDP is typically defined by the tuple $\langle \mathcal{S}, \mathcal{A}, \mathcal{P}, \mathcal{R}, \gamma \rangle$, where:

\begin{itemize}
    \item \textbf{State Space} $\mathcal{S}$: The set of all possible environment states. A state $s_t \in \mathcal{S}$ at time $t$ satisfies the Markov property, meaning it contains all relevant information from the history necessary to make future decisions.
    
    \item \textbf{Action Space} $\mathcal{A}$: The set of all available actions the controller can take.
    
    \item \textbf{Transition Function} $\mathcal{P}$: The probability distribution $\mathcal{P}(s_{t+1} | s_t, a_t)$ which gives the likelihood of transitioning to state $s_{t+1}$ after taking action $a_t$ in state $s_t$.
    
    \item \textbf{Reward Function} $\mathcal{R}$: A scalar reward $r_{t+1} = \mathcal{R}(s_t, a_t)$ received by the controller after taking action $a_t$ in state $s_t$ (equivalently, we write $r_t$ in later sections where the subscript indexes the decision step rather than the post-transition tick). The primary objective of the controller is to maximize the expected return, defined as:
    \begin{equation}
        \mathcal{G}_t = \sum_{k=0}^{\infty} \gamma^k r_{t+k+1},
    \end{equation}
    where $\gamma \in [0, 1)$ is the discount factor.
    
    \item \textbf{Discount Factor} $\gamma$: A parameter that controls the trade-off between immediate and future rewards. A value of $\gamma$ close to $0$ makes the controller ``myopic'' (focused on immediate rewards), while a value close to $1$ prioritizes long-term gains, which is essential for continuing tasks.
\end{itemize}

The controller's behavior is determined by a policy $\pi(a|s)$, which represents the probability of selecting action $a$ given state $s$. The optimal policy maximizes the expected return. 

\noindent
The subsequent sections will present the formulation of TSC as a RL problem, and its adaptation to LLHE traffic.

\section{Problem Formulation}\label{sec:formulation}
The TSC problem is formulated as a MDP in which, at each discrete time step $t$, the controller observes the current traffic state, selects an action to adjust signal timings, and receives a reward reflecting the resulting congestion level. The components of this MDP formulation are defined below.

\begin{figure*}[t]
\centering
\includegraphics[width=0.7\linewidth]{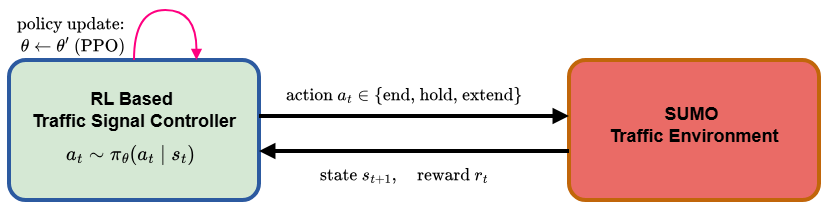}

\caption{Reinforcement learning loop for the proposed traffic signal controller.
At each decision step, the controller observes state $s_t$ (active phase, cycle
timing, queue lengths) and selects a discrete action
$a_t \in \{\text{end}, \text{hold}, \text{extend}\}$ to adjust the current
phase's green time. The SUMO environment simulates the resulting traffic and
returns the next state $s_{t+1}$ and queue based reward $r_t$
(Eq.~\eqref{eq:reward}). Experience tuples $(s_t,a_t,r_t,s_{t+1})$ update the
policy $\pi_\theta$ via the PPO clipped objective (Section~\ref{sec:ppo}).}
\label{fig:methodology}
\end{figure*}

\subsection{State Space $\mathcal{S}$}
The state space represents a snapshot of the dynamic environment that the RL controller observes to select an informed action, e.g., adjusting the green time of the current phase to minimize congestion. At each decision step $t$, the canonical 8-D state is defined as
\begingroup\color{black}
\begin{equation}
s_t = \left( T_c,\; P,\; t_p,\; Q,\; g_1,\, g_2,\, g_3,\, g_4 \right),
\label{eq:state}
\end{equation}
\endgroup
\textcolor{black}{where $T_c$ is the current cycle length, $P$ is the (scalar) index of the active phase, $t_p$ is the time remaining in the current phase, $Q$ is the total queue length across all phases, and $g_i$ denotes the green time of phase $i$ for $i = 1,\dots,N_p$ ($N_p$ is the total number of phases, set to 4 for lane-based site). This yields 8 scalar components.} This is referred to as the \textbf{raw-queue representation} (reward-labelled \textbf{PPO (Queue)}) throughout this paper. The expanded 19-D state defined in Section~\ref{sec:State}, denoted $s_t^{(19)}$, reuses several of these symbols (e.g.\ $T_c$, $t_p$, $P$, $Q$) with normalised and vector-valued meanings that are stated explicitly there; the two states should not be confused.

Rather than conditioning the policy directly on $s_t$, we additionally experiment with a representation function $\phi$ that maps $s_t$ (or an expanded variant of it) to an observation $z_t = \phi(s_t) \in \mathbb{R}^d$ consumed by the PPO policy (Fig.~\ref{fig:state-interface}). 

\subsection{Action Space $\mathcal{A}$}
At every $\delta_{\text{time}}$ seconds (set to 5 seconds), the RL controller selects a discrete action $a_t$ from a predefined set to modify the green time allocation of the current traffic signal phase. The action space is defined as: $a_t \in \{0, 1, 2\}$, the actions and their physical interpretation are as follows:\newline

\resizebox{\columnwidth}{!}{%
\begin{tabular}{ll}
\textbf{0} & End the current phase immediately. \\
\textbf{1} & Continue the current phase with no change \\ 
& (i.e., maintain green time for another $\delta_{\text{time}}$ seconds). \\
\textbf{2} & Extend the current phase by an additional $\delta_{\text{time}}$ seconds. \\
\end{tabular}%
}

\subsection{Reward Function $\mathcal{R}$}
\label{subsec:reward_function}
The reward function is designed to reflect the level of congestion at the signalized intersection, based on the observation that queue lengths are a direct and measurable indicator of congestion. 

The reward combines two objectives: penalizing the current absolute queue size and rewarding progress in queue reduction.
A vehicle is considered part of a queue if its speed is less than 0.1 m/s. For each approach $j \in \{1, \dots, N_{\text{app}}\}$ (e.g., North, East, South, West), let $\mathcal{L}_j$ be the set of lanes in that approach and  $q_{l,t}$ denote the queue length of an individual lane $l$ at timestep $t$. The queue length for the approach  $Q_{j,t}$, is expressed as the maximum queue length observed across any of its lanes at that timestep:

\begin{equation}
Q_{j,t} = \max_{l \in \mathcal{L}_j} q_{l,t}.
\end{equation}

The final reward $r_t$ is a weighted sum that combines a penalty for the absolute queue size with a reward for reducing the queue relative to the previous timestep:

\begin{equation}
r_t = \alpha_{abs} \left( - \frac{\sum_{j=1}^{N_{\text{app}}} Q_{j,t}}{N_{\text{app}} \cdot Q_{\text{norm}}} \right) + \alpha_{red} \left( \frac{\sum_{j=1}^{N_{\text{app}}} (Q_{j, t-1} - Q_{j,t})}{N_{\text{app}} \cdot Q_{\text{norm}}} \right),
\label{eq:reward}
\end{equation}

where the weights $\alpha_{abs}, \alpha_{red}$ were set to $(0.4, 0.6)$ respectively (results for other weightings are listed for reference in Appendix~\ref{app:alpha}). $N_{\text{app}}$ is the number of approaches and $Q_{\text{norm}}$ is a normalization constant. 

Eq.~\eqref{eq:reward} combines two complementary signals: an absolute-congestion term that penalizes standing queue length, and a queue-reduction term that credits the trend in queue length between consecutive steps. This lets the controller be rewarded for actively clearing a queue even while it remains, in absolute terms, the largest at the intersection -- a case a purely absolute formulation cannot distinguish from a static or growing queue. 

\subsection{Operational Constraints}
For the study, a 4-phase signalized intersection is used, with green times bounded by $g_{\min} \leq g_i \leq g_{\max}$ for phase $i$: $g_{\min} \approx 10\,\text{s}$, following the minimum pedestrian crossing time \cite{manual2000highway}, and $g_{\max} = 40\,\text{s}$, following the codal provision [IRC SP: 41-1994] \cite{congress1994guidelines}. Actions are only available once the active phase's green time exceeds this 10-second minimum; before that threshold the phase continues by default. The minimum yellow time is separately fixed at $Z_{\min} \approx 5\,\text{s}$ using the standard dilemma-zone model \cite{gartner2002traffic}. Both derivations follow standard traffic-engineering procedure and are given in full in Appendix~\ref{app:constraints}.

\subsection{Extension to LLHE Traffic}
\label{sec:hetero-formulation}

The MDP formulation is unchanged for LLHE Traffic, except that queues are measured per \emph{approach} rather than per lane: $\mathcal{L}_j$ becomes the set of lanes or edges carrying approach $j$ traffic, and $q_{l,t}$ is measured over the full approach width (Section~\ref{sec:laneless-sumo}). The reward, action space, and green-time bounds are retained as defined.

\section{Proposed Methodology}\label{sec:method}
The overall methodology is illustrated in Fig.~\ref{fig:methodology}. A PPO based controller interacts with the SUMO traffic simulator \cite{krajzewicz2010traffic} (or the environment): at each decision step it observes the current traffic state, selects an action that adjusts the green times within the ongoing cycle, and receives a queue based reward. Over successive iterations, the controller learns a policy that dynamically adjusts signal timings to minimize congestion. When applied to the LLHE site, the SUMO environment additionally simulates a heterogeneous vehicle mix without enforced lane discipline (more details in Section~\ref{sec:laneless-sumo}). 

\begin{figure}[t]
\centering
\resizebox{\columnwidth}{!}{%
\begin{tikzpicture}[
  node distance=6mm and 8mm,
  box/.style={draw, rounded corners, align=center, minimum height=9mm, minimum width=20mm, font=\scriptsize},
  arr/.style={-{Latex[length=2mm]}, thick}
]
\node[box] (state) {Raw state $s_t$\\(19-D)};
\node[box, right=of state] (enc) {Autoencoder\\encoder $E$};
\node[box, right=of enc] (latent) {Latent obs.\\$z_t$ ($d$-D)};
\node[box, right=of latent] (policy) {PPO policy\\$\pi(a \mid z_t; \theta)$};
\node[box, below=of state, minimum width=20mm] (identity) {Identity\\(no compression)};

\draw[arr] (state) -- (enc);
\draw[arr] (enc) -- (latent);
\draw[arr] (latent) -- (policy);
\draw[arr] (state.south) -- (identity.north);
\draw[arr] (identity.east) -| (policy.south);
\end{tikzpicture}%
}
\caption{State-representation interface: the raw 19-D state is optionally passed through the autoencoder encoder $E$ to produce the latent observation $z_t$ consumed by the PPO policy; the same interface is reused across the 4/8/16/19/32-D autoencoder configurations by varying only $d$. The Full-State and raw-Queue configurations bypass $E$ and feed the (sub-set of the) raw state directly to the policy, shown as the lower path.}
\label{fig:state-interface}
\end{figure}

\vspace{-1em}

\subsection{State Representation Designs} \label{sec:State}

This section details the family of representation functions $\phi$ that map the raw traffic state $s_t$ (or an expanded variant of it) to the observation $z_t=\phi(s_t)\in\mathbb{R}^d$ consumed by the PPO policy (see Section~\ref{sec:formulation}), aiming to identify a formulation that is data-efficient yet expressive. Table~\ref{tab:representations} summarizes the representations implemented for the LBHO site. The baseline 8-D raw-queue state of Eq.~\eqref{eq:state} serves as the reference point; the 19-D expansion introduced below is called \textbf{Full State} (reward-labelled \textbf{PPO (Full State)}), a name used consistently from here onward. Throughout this paper, controllers are labelled by the pair (reward, state): the raw-queue 8-D state trained under the queue based reward is written PPO (Queue), the Full State trained under the queue based reward is written PPO (Full State), and an autoencoder (AE) with latent size $d$ trained under the queue based reward is written PPO (Queue AE-$d$D). Where the Delay, Pressure or Speed rewards are used, the state is the 19-D Full State unless noted otherwise.

\begin{table}[h!]
\centering
\caption{Overview of different state representations (LBHO site)}
\label{tab:representations}
\resizebox{\columnwidth}{!}{%
\begin{tabular}{|l|l|l|l|}
\hline
\textbf{Representation} & \textbf{Dimensions} & \textbf{Key Features} & \textbf{Motivation} \\
\hline
PPO (Queue)            & 8-D  & Basic traffic state   & Minimal complexity \\
\hline
PPO (Full State)       & 19-D & Detailed per-lane info & Capture spatial patterns \\
\hline
Autoencoder            & Variable & Learned compression    & \textcolor{black}{Lower input dimensionality} \\
\hline
\end{tabular}%
}
\end{table}

The baseline 8-D representation ($\phi = \mathrm{id}$, i.e.\ $z_t = s_t^{(8)}$) is precisely the state of Eq.~\eqref{eq:state}, reported as ``8-D'' in Table~\ref{tab:representations}.

\subsubsection{Expanded Representation (19-D)}
The expanded state representation $s_t^{(19)}$ is designed to provide richer information for a better understanding of traffic conditions:
\begin{equation}
s_t^{(19)} = \bigl[T_c,\; \mathbf{P},\; t_p,\; N_{\text{cycles}},\; \mathbf{Q},\; \Delta\mathbf{Q},\; \mathbf{g}\bigr],
\label{eq:state_full}
\end{equation}
where the symbols are re-used from Eq.~\eqref{eq:state} with the modified, normalised meanings listed below (note in particular that $T_c$ and $t_p$ are elapsed rather than remaining times, $\mathbf{P}$ is a one-hot vector rather than a scalar index, and $\mathbf{Q}$ is a per-approach vector rather than a scalar total):

\begin{itemize}
  \item $T_c$: Normalized time elapsed in the current cycle.
  \item $\mathbf{P}$: A 4-D one-hot encoding of the active green phase.
  \item $t_p$: Normalized time elapsed in the current phase.
  \item $N_{\text{cycles}}$: Normalized count of completed signal cycles.
  \item $\mathbf{Q}$: A 4-D vector of normalized maximum queue lengths per cardinal approach, $Q_j = \text{queue}_j / \text{queue}_{\max}$, where $\text{queue}_{\max}$ coincides with the reward normaliser $Q_{\text{norm}}$ of Eq.~\eqref{eq:reward}.
  \item $\Delta\mathbf{Q}$: A 4-D vector of normalized queue-length changes, $\Delta Q_j = (\text{queue}_{j,t} - \text{queue}_{j,t-1})/\text{queue}_{\max}$. By this sign convention, a \emph{positive} $\Delta Q_j$ indicates a growing queue, so the queue-reduction term in Eq.~\eqref{eq:reward} (which uses $Q_{j,t-1}-Q_{j,t}$) carries the opposite sign.
  \item $\mathbf{g}$: A 4-D vector of normalized programmed green times, $g_i = \text{green\_time}_i / \text{green\_time}_{\max}$.
\end{itemize}

\subsubsection{Learned Representations via Autoencoder}
An autoencoder (AE) learns a latent representation $z_t \in \mathbb{R}^d$ of the 19-D state $s_t^{(19)}$ (architecture in Table~\ref{tab:ae_architecture}). Here an encoder $E$ maps the state to the latent space and a decoder $D$ reconstructs it:
\begin{equation}
z_t = E(s_t^{(19)};\,\theta_E), \qquad \hat{s} = D(z_t;\,\theta_D).
\end{equation}
Both are trained jointly and offline by minimizing the mean-squared reconstruction loss over a buffer $\mathcal{D}$ of states collected under random-action rollouts:
\begin{equation}
\min_{\theta_E,\,\theta_D}\;\mathcal{L} = \frac{1}{|\mathcal{D}|} \sum_{s \in \mathcal{D}} \|s - \hat{s}\|^2_2.
\end{equation}
The encoder is then frozen and the PPO policy receives $z_t$ as its observation. The latent dimension is swept over $\{4, 8, 16, 19, 32\}$, covering compressive ($d<19$), identity-sized ($d=19$) and overcomplete ($d=32$) settings.

The raw-queue representation at the LLHE site follows the same 8-D structure as Table~\ref{tab:representations}, with $Q$ and the per-phase green times re-interpreted at approach level per Section~\ref{sec:hetero-formulation}. The AE family was retrained on the 19-D LLHE-site state with the same latent sizes and objective.

\begin{table}[h!]
\centering
\scriptsize
\caption{Autoencoder architecture and training configuration, shared
across all latent sizes $d \in \{4, 8, 16, 19, 32\}$ and both study
sites.}
\label{tab:ae_architecture}
\begin{tabular}{|l|l|}
\hline
\textbf{Parameter} & \textbf{Value} \\
\hline
Encoder layer widths       & 19--64--128--$d$ \\
Decoder layer widths       & $d$--128--64--19 \\
Activation function        & ReLU (linear output) \\
Optimizer                  & Adam \\
Learning rate              & $10^{-3}$ \\
Batch size                 & 256 \\
Training epochs            & 30 \\
State buffer size $|\mathcal{D}|$ & 10{,}000 (random-action rollouts) \\
Early-stopping criterion   & None \\
\hline
\end{tabular}
\end{table}

\subsection{Policy Optimization}
\label{sec:ppo}
The controller is trained with Proximal Policy Optimization (PPO)~\cite{schulman2017proximal}, a policy-gradient method whose clipped surrogate objective limits the change between successive policies. As in Fig.~\ref{fig:methodology}, at each decision step the policy $\pi_\theta$ receives the observation $z_t$ (the raw state or an AE encoding), selects $a_t$, and SUMO returns $r_t$ and the next state. The policy is updated from the collected rollouts for a fixed budget of 100{,}000 decision steps. Implementation and hyperparameters are given in Section~\ref{sec:setup} and Table~\ref{tab:all_hyperparams}.

\subsection{Training Framework}
The training procedure follows the standard on-policy loop illustrated in Fig.~\ref{fig:methodology}. At each decision point the controller observes state $s(t)$, selects action $a(t)$ according to its policy $\pi$, and the SUMO environment returns a reward $r(t)$ and the next state $s(t{+}1)$. The PPO algorithm collects these experience tuples over multiple steps and uses them to update the policy. This cycle is repeated until convergence.
\begin{figure*}[t]
\centering
\subfloat[\footnotesize LBHO study site]{%
\includegraphics[width=0.45\linewidth]{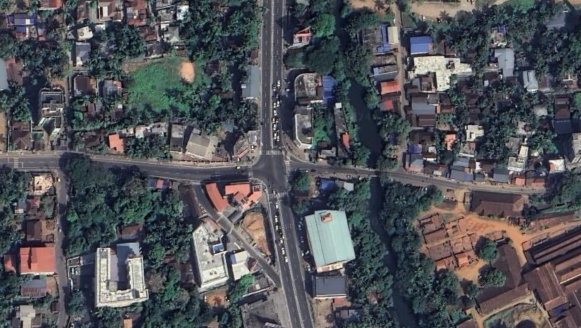}%
\label{fig:site-figure}%
}\hspace{0.05\textwidth}
\subfloat[\footnotesize LBHO SUMO simulation]{%
\includegraphics[width=0.45\linewidth]{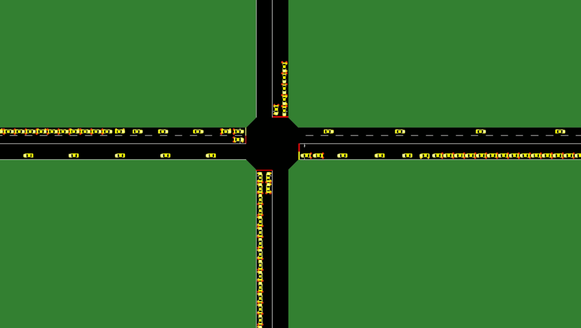}%
\label{fig:simulation}%
}
\vspace{0.5em}
\subfloat[\footnotesize LLHE study site]{%
\includegraphics[width=0.45\linewidth]{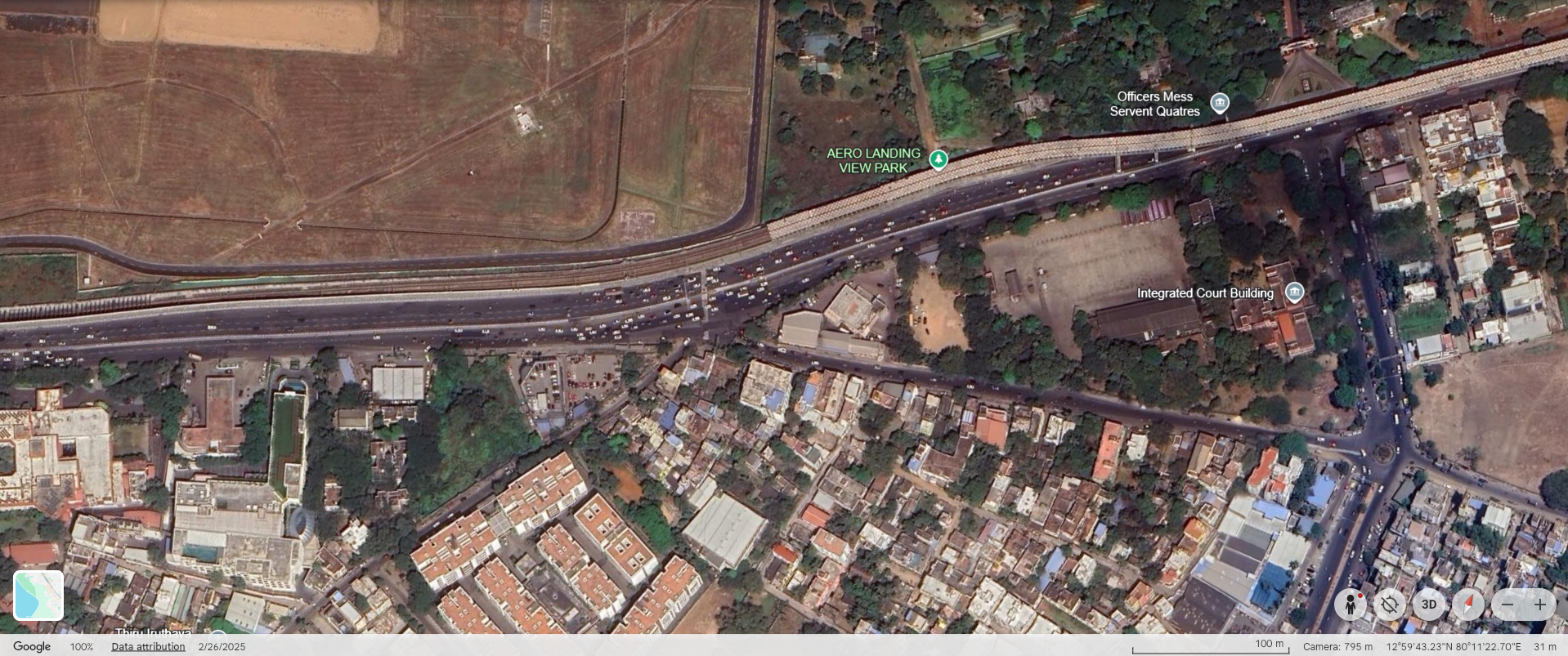}%
\label{fig:hetero-site}%
}\hspace{0.05\textwidth}
\subfloat[\footnotesize LLHE SUMO simulation]{%
\includegraphics[width=0.45\linewidth]{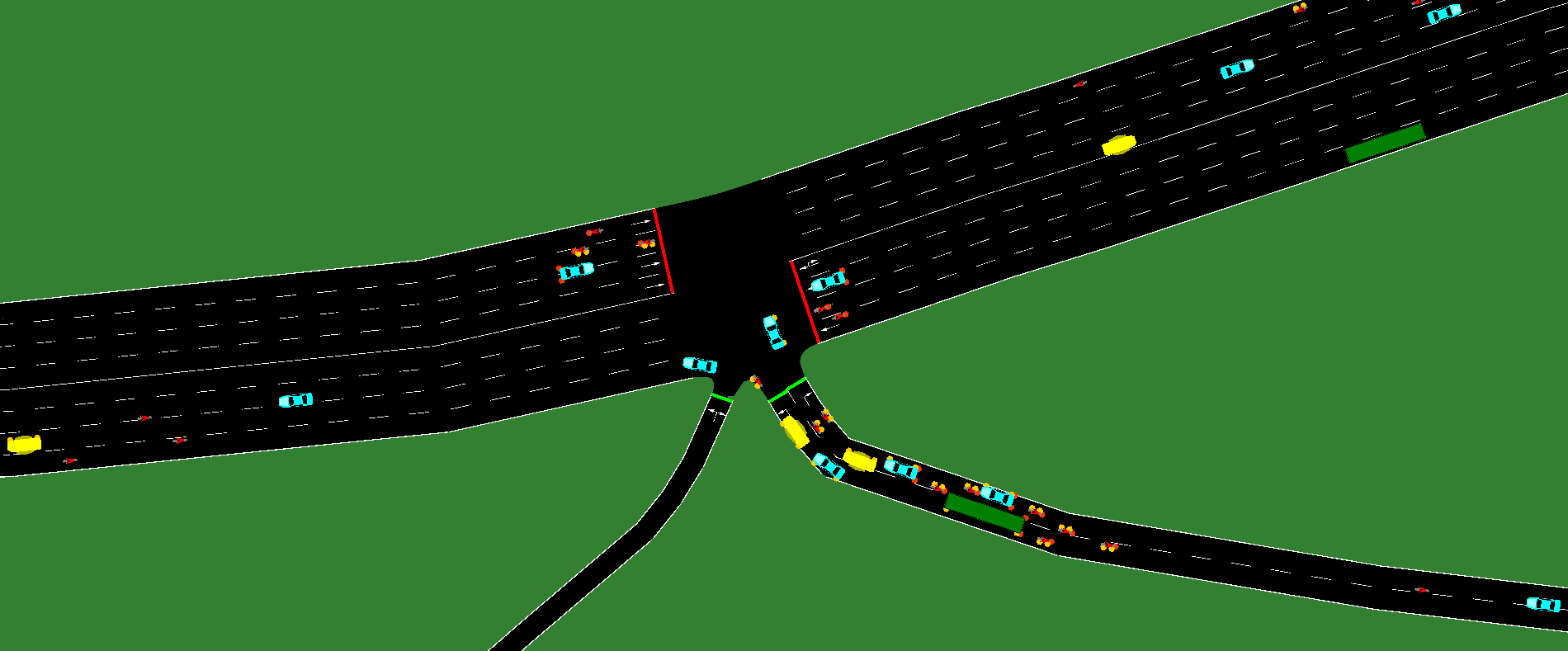}%
\label{fig:hetero-sim}%
}
\caption{Study sites and their SUMO simulation environments. \textbf{(a)-(b)} LBHO site. \textbf{(c)-(d)} LLHE site; the junction is an unbalanced layout where two major bidirectional approaches meet minor feeder approaches at oblique angles.}
\label{fig:all_sites}
\end{figure*}

\section{Simulating LLHE Traffic in SUMO} \label{sec:laneless-sumo} 
SUMO's conventional configuration assumes strict lane discipline: each edge is divided into a fixed number of parallel lanes of a fixed width, vehicles are assigned to a single lane at a time, car-following is computed along that lane's centreline, and lane changes are treated as discrete, instantaneous transfers between adjacent lanes. 
However this assumption doesnt apply for sites or scenatios where diverse vehicles share the same road width without confining themselves to marked lanes. While prior work has explored adapting SUMO for laneless, heterogeneous traffic \cite{patel2016comparison}, these efforts target older SUMO versions and do not include an RL TSC integration. 

To our knowledge, ours is the first openly available implementation to combine a LLHE SUMO configuration on a current SUMO version with reinforcement learning for signal control. 

\subsection{Heterogeneous Vehicle Composition}
Rather than a single vehicle type, the simulated traffic stream is composed of multiple vehicle classes representative of the study site: two-wheelers, private cars, auto-rickshaws (three-wheelers) and heavy motor vehicles (buses and other commercial vehicles, simulated as one class). Each class is assigned its own physical length, width, maximum speed, maximum acceleration, and comfortable deceleration, together with a class-specific proportion of the overall traffic mix drawn from the field-collected flow data. The resulting composition is summarised in Table~\ref{tab:veh_composition}.
\begin{table}[h]
\centering
\caption{Heterogeneous vehicle composition derived from field-collected flow data.}
\label{tab:veh_composition}
\begin{tabular}{|l|l|}
\hline
\textbf{Vehicle Class} & \textbf{Share (\%)} \\
\hline
Two-wheeler & 64.0 \\
\hline
Car & 25.7 \\
\hline
Auto-rickshaw & 6.3 \\
\hline
Heavy Motor Vehicle & 3.9 \\
\hline
\end{tabular}
\end{table}

\subsection{Laneless Lateral Movement}
SUMO's default lane-changing model treats lateral movement as a discrete, instantaneous transfer between adjacent lane centrelines. To reproduce the continuous lateral filtering observed at the study site, all vehicle types are assigned the \texttt{SL2015} sublane model, which tracks a continuous lateral offset at sub-lane resolution and replaces discrete lane changes with gradual lateral adjustments. 

The key parameters that differ from SUMO's defaults are the lateral alignment preference, which removes the default bias towards a lane's centreline and allows vehicles to occupy any lateral position, and the minimum lateral gap, which permits the close side-by-side positioning typical of mixed traffic rather than enforcing a full-lane separation. Together, these settings allow vehicles of all classes to filter, overtake, and queue side by side across the carriageway width while retaining the same signal-phase and stop-line logic as the baseline configuration. A summary of the key parameter differences is given in Table~\ref{tab:lateral_params}.
\begin{table}[h]
\centering
\caption{Lateral movement parameters: SUMO defaults vs.\ this configuration.}
\label{tab:lateral_params}
\begin{tabular}{|p{2.2cm}|p{2.6cm}|p{2.6cm}|}
\hline
\textbf{Parameter} & \textbf{SUMO Default} & \textbf{This Work} \\
\hline
Lane-change model & \texttt{LC2013} (discrete) & \texttt{SL2015} (sublane) \\
\hline
Lateral alignment & \texttt{center} & \texttt{arbitrary} \\
\hline
Min.\ lateral gap (\texttt{minGapLat}) & 0.6\,m & 0.5\,m \\ Lateral resolution & none (no sublanes) & 0.8\,m \\
\hline
Lateral tracking & Implicit (lane ID) & Continuous offset \\
\hline
Handedness & Right-hand & Left-hand \\
\hline
\end{tabular}
\end{table}

\subsection{Approach-Level Traffic Descriptors}
At the LLHE site, since individual lanes are no longer a meaningful unit of observation once vehicles are free to take up any lateral position, all traffic descriptors used by the RL controller and by the evaluation metrics, including queue length, flow, and waiting time, are aggregated at the level of the whole approach rather than a single marked lane, consistent with the redefinition of $\mathcal{L}_j$ given in Section \ref{sec:hetero-formulation}. This mirrors how queue length would in practice be estimated at an LLHE intersection using an overhead camera or similar approach-level sensor, rather than lane-specific detector loops.

\section{Experimental Setup}\label{sec:setup}
The framework is evaluated at two sites that represent contrasting traffic regimes. The LBHO site, a conventional lane-based intersection with homogeneous traffic, is the setting in which RL-based TSC has mostly been developed, so it serves as a reference for validating the framework. The LLHE site, a laneless junction with heterogeneous traffic, is the target setting that motivates the queue-based reward. This section describes both sites, their simulation in SUMO \cite{krajzewicz2010traffic}, the implementation details, and the evaluation metrics.
\vspace{-1em}
\subsection{Study Site: LBHO Traffic}
This study focuses on an intersection on Puthiyara Road in Kozhikode, a major thoroughfare characterized by high traffic volume, as shown in Fig. \ref{fig:site-figure}. This area is a common connection point for various types of vehicles, including private cars, commercial vehicles, public buses, and taxis. Due to the increasing number of vehicles and the road's capacity, the area is prone to traffic congestion, particularly during peak hours.
\vspace{-1em}

\begin{figure}[htbp]
\centering
\subfloat[\footnotesize Lane-Based Movement]{%
\includegraphics[width=0.6\linewidth]{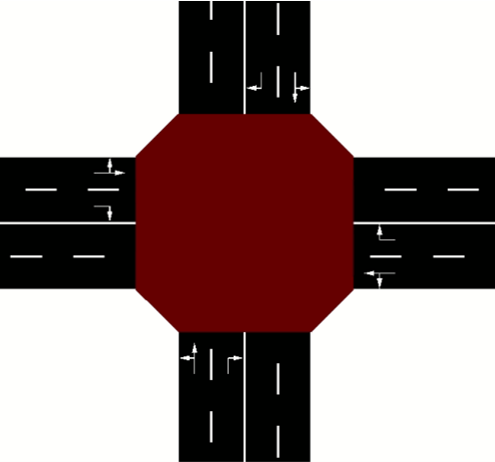}%
\label{fig:lanebased_movement}
}\\
\subfloat[\footnotesize Phase Diagram]{%
\includegraphics[width=0.9\linewidth]{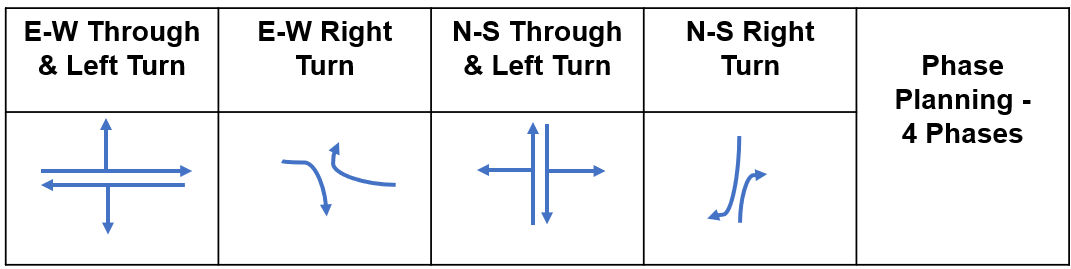}%
\label{fig:phase_diagram}
}
\caption{Illustration of the LBHO intersection configuration: (a) lane-based movement and (b) corresponding signal phasing diagram.}
\label{fig:intersection_and_phasing}
\end{figure}
\subsection{SUMO Environment: LBHO Site}
\begin{figure}[htbp]
\centering
\includegraphics[width=0.48\textwidth]{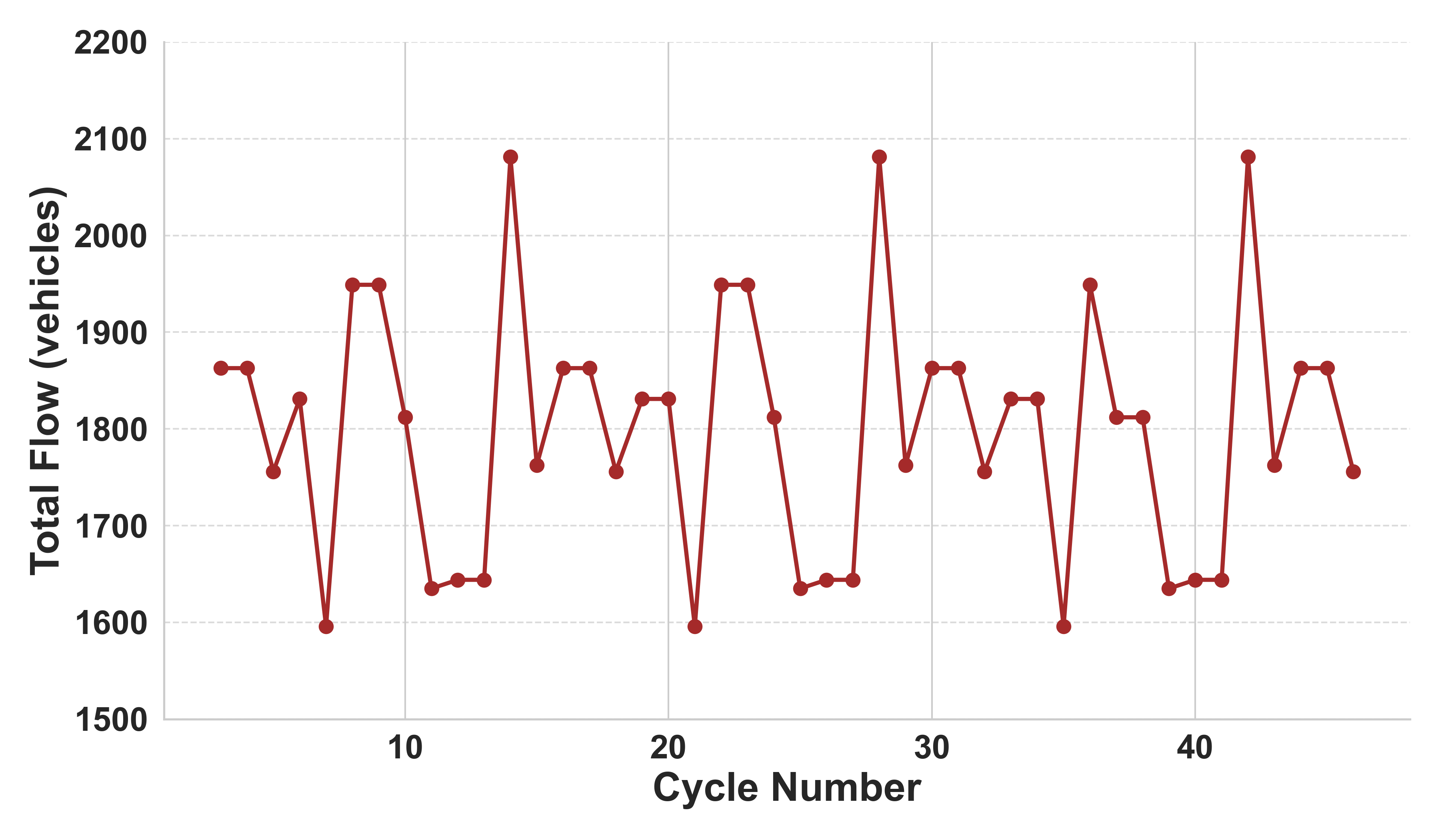}
\caption{Total flow of vehicles from the field data collected at the LBHO site.}
\label{fig:flow}
\end{figure}

Fig. \ref{fig:simulation} shows the study site generated using SUMO traffic simulator \cite{krajzewicz2010traffic}. The simulation incorporates a 4-phase signalized intersection with lane-based movements under left-hand traffic conditions as shown in Fig. \ref{fig:lanebased_movement}. It assumes homogeneous traffic and replicates flow patterns based on field data collected from the study site (Fig.~\ref{fig:flow}). Each approach has two lanes, one for through and left movement and the other for right movement, with a lane width of 3.1 meters. It is a four phase signal with details as shown in Fig. \ref{fig:phase_diagram}.
\vspace{-1em}

\subsection{Study Site: LLHE Traffic}
The LLHE site of this study is a signalized junction near Chennai International Airport, a location that experiences heavy traffic volume of varied vehicle classes such as cars, two-wheelers, auto-rickshaws, commercial vehicles, public buses, etc. Unlike the LBHO site, no lane discipline was found here, and hence vehicles of different classes were assumed to occupy any position across the approach width, consistent with the laneless driving behaviour observed at this location. Geometrically, the junction is an unbalanced, non-orthogonal layout in which a West approach, a Northeast approach, a Southeast approach, and a Southwest approach meet at oblique angles, reflecting a real merge-and-through junction rather than a textbook four-way intersection as shown in Fig.~\ref{fig:hetero-site}.

\subsection{SUMO Environment: LLHE Site}
The LLHE study site was likewise generated using the SUMO traffic simulator \cite{krajzewicz2010traffic} (Fig.~\ref{fig:hetero-sim}), using the laneless, heterogeneous-vehicle configuration described in Section \ref{sec:laneless-sumo}. The simulation incorporates a multi-phase signalized intersection with approach based, laneless movements under left-hand traffic conditions. It assumes a heterogeneous vehicle mix without enforced lane discipline, and replicates flow patterns based on field data collected from the study site.

\subsection{Implementation Details}
PPO was implemented for the TSC task using the \textit{stable-baselines3} \cite{raffin2021stable} package and trained for 100,000 decision steps ($\delta_{\text{time}}=5$\,s, i.e.\ $\approx$500,000\,s simulated) at both sites. Hyperparameters were tuned independently for each site; the LLHE site required a larger batch size and step count to stabilise training, as initial runs with default settings showed high variance across the five random seeds. The final hyperparameters are given in Table~\ref{tab:all_hyperparams}. All runs used an NVIDIA RTX A5000 GPU; the training took approximately 1.1-1.7 hours per run, and autoencoders for state abstraction were trained separately prior to policy training. Simulations were run in SUMO 1.22.0.

\begin{table}[htbp]
\centering
\caption{PPO training hyperparameters and budget, LBHO and LLHE sites.}
\label{tab:all_hyperparams}
\resizebox{\columnwidth}{!}{%
\begin{tabular}{lcc}
\toprule
\textbf{Hyperparameter} & \textbf{LBHO} & \textbf{LLHE} \\
\midrule
Learning Rate              & $3\times10^{-6}$ & $3\times10^{-4}\to1\times10^{-5}$ \\
Learning-Rate Schedule     & constant & linear decay \\
Optimizer                  & Adam & Adam \\
Batch Size                 & 64 & 256 \\
Number of Steps            & 200 & 2048 \\
Number of Epochs           & 10 & 5 \\
Rollout Buffer Size        & 200 transitions & 2048 transitions \\
Discount Factor            & 0.99 & 0.99 \\
GAE Weight                 & 0.95 & 0.95 \\
Clip Range                 & 0.2 & 0.2 \\
Entropy Coefficient        & 0 & 0.01 \\
Value-Function Coefficient & 0.5 & 0.5 \\
Max Gradient Norm          & 0.5 & 0.5 \\
Advantage Normalisation    & yes & yes \\
Policy / Value Network     & MLP $64\text{--}64$ (tanh) & MLP $64\text{--}64$ (tanh) \\
Training Episode Length    & 105{,}000\,s & 1{,}800\,s \\
Total Training Timesteps   & 100{,}000 steps & 100{,}000 steps \\
Number of Seeds            & 5 & 5 \\
\bottomrule
\end{tabular}%
}
\end{table}

\subsection{Evaluation Metrics}

The total queue length per cycle is used as the primary evaluation metric to assess the performance of the traffic signal controller at both sites. For each approach in the intersection (e.g., North, South, East, West at the LBHO site; West, Northeast, Southeast, Southwest at the LLHE site), the maximum queue length is computed across all lanes, or across the full approach width at the LLHE site, during the cycle. Let $j \in \{1, \dots, N_{\text{app}}\}$ denote an approach at the intersection. For each approach $j$, let $\mathcal{L}_j$ be the set of lanes belonging to approach $j$, and for each lane $l \in \mathcal{L}_j$, let $q_l^{\max}$ represent the maximum queue length observed on lane $l$ during the cycle. The total queue length for approach $j$ is computed as the maximum queue length across its lanes:
\begin{equation}
Q^{\text{approach}}_j = \max_{l \in \mathcal{L}_j} q_l^{\max},
\label{eq:approach-queue}
\end{equation}
where $Q^{\text{approach}}_j$ is the maximum queue observed in the entire approach $j$ during the cycle. The total queue length for the entire cycle is then computed as the sum of maximum queue lengths from all approaches:
\begin{equation}
Q_{\text{cycle}} = \sum_{j=1}^{N_{\text{app}}} Q^{\text{approach}}_j,
\label{eq:cycle-queue-approach}
\end{equation}
where $Q_{\text{cycle}}$ represents the total queue length for the cycle based on per-approach maximum queues. Queue values are reported in the units natively returned by SUMO's \textit{getLastStepHaltingNumber} (number of halted vehicles per approach). Mean waiting time (seconds) and mean speed (m/s) are also recorded, primarily on the LLHE site and also on the LBHO site for the build-up study (Table~\ref{tab:dim_buildup}); both are natively output by SUMO and are widely used in prior RL-TSC works such as \cite{noaeen2022reinforcement, xiao2026intelligent}. 

Since the LBHO site aggregates queues as a max over marked lanes while the LLHE site aggregates them over the full approach width, queue values are not directly comparable across the two sites. \textcolor{black}{The cycle length is itself set by the controller, so the window over which the per-cycle maximum is taken differs between controllers; a controller that shortens its cycles takes the maximum over a shorter window, which biases this metric downward. This caveat applies to all queue comparisons in Section~\ref{sec:results}.}

\section{Baseline Methods}\label{sec:baselines}
The proposed controller is compared against two non-learning baselines (Dynamic Webster and SUMO's built-in actuated control), a learned RL baseline (RESCO-DQN), and PPO controllers trained with alternative rewards.

\vspace{-1em}
\subsection{Dynamic Webster Method}
Webster's method~\cite{webster1958traffic} computes signal timings from observed traffic flows. The optimum cycle length $C_o$ is
\begin{equation}
C_o = \frac{1.5L_{\text{lost}} + 5}{1 - \sum_{i=1}^{N_p} Y_i},
\end{equation}
where $L_{\text{lost}}$ is the total lost time per cycle (s), $N_p$ is the number of signal phases, and $Y_i$ is the maximum flow ratio $q_{ij}/S_j$ over the lane groups $j$ served by phase $i$, with $q_{ij}$ the flow and $S_j$ the saturation flow of lane group $j$. To make the comparison with adaptive control fair, the timings are recomputed dynamically from updated flow estimates every 145\,s. At the LBHO site, the flow estimates are additionally moving averages over the past 15 minutes.

\vspace{-1em}
\subsection{Actuated Control (SUMO)}
The second non-learning baseline is SUMO's built-in actuated controller~\cite{krajzewicz2010traffic}, which extends or terminates each green phase based on vehicle gaps measured by induction-loop detectors, within minimum and maximum green bounds.

\vspace{-1em}
\subsection{RESCO-DQN}
RESCO~\cite{NEURIPSDATASETSANDBENCHMARKS2021_f0935e4c} is a toolkit for developing and comparing RL-based traffic signal controllers on realistic scenarios, and reports that DQN outperforms many previously reported state-of-the-art methods. 
Following this, a DQN controller with independent learners per intersection was implemented. Its state is a matrix of per-lane features (active phase indicator, approaching vehicle count, total wait time, queue length, and sum of vehicle speeds), which convolutional layers aggregate across lanes.

At the LLHE site, the per-lane state matrix is built from approach-level aggregates, consistent with Section~\ref{sec:hetero-formulation}. The reward is the negative sum of waiting times across all lanes, normalized and clipped for stability:
\begin{equation}
R_t = \text{clip}\left(\frac{-\sum_{l \in \text{Lanes}} \text{wait}_l}{\alpha}, R_{\min}, R_{\max}\right),
\end{equation}
where $\text{wait}_l$ is the total wait time in lane $l$, $\alpha$ is a normalization factor, and $R_{\min}$, $R_{\max}$ are the reward clipping limits.

\vspace{-1em}
\subsection{Alternate Reward Formulations}
\label{subsec:alt-rewards}
To assess the proposed queue based reward, PPO controllers were also trained with the delay, pressure and average-speed rewards defined below.

\subsubsection{Delay based Reward}
This reward credits the reduction in average accumulated delay across all lanes between consecutive time steps:
\begin{equation}
R_{\text{wait}}(t) = \mathcal{W}(t-1) - \mathcal{W}(t),
\end{equation}
where $\mathcal{W}(t)$ is the average accumulated delay across all lanes at time $t$:
\begin{equation}
\mathcal{W}(t) = \frac{1}{N_{\text{lane}}} \sum_{i=1}^{N_{\text{lane}}} w_i(t),
\end{equation}
$w_i(t)$ is the accumulated delay on lane $i$, and $N_{\text{lane}}$ is the total number of lanes. (The symbol $\mathcal{W}$ is distinct from the intersection width $W$ used in Appendix~\ref{app:constraints}.)

\subsubsection{Pressure based Reward}
Pressure is the difference between the inflow and outflow of vehicles on each lane, which indicates build-up or relief of congestion. The reward is the negative of the total pressure, so \emph{maximising} it balances arrivals and departures at the intersection:
\begin{equation}
R_{\text{pressure}}(t) = -\sum_{i=1}^{N_{\text{lane}}} \left( q^{\text{in}}_i(t) - q^{\text{out}}_i(t) \right),
\end{equation}
where $q^{\text{in}}_i(t)$ and $q^{\text{out}}_i(t)$ are the numbers of vehicles approaching and leaving lane $i$ at time $t$, and $N_{\text{lane}}$ is the number of lanes. This is an instantaneous measure.

\begin{figure*}[t]
\centering
\subfloat[\footnotesize Total Queue Length vs Cycle Number]{%
\includegraphics[width=0.48\linewidth]{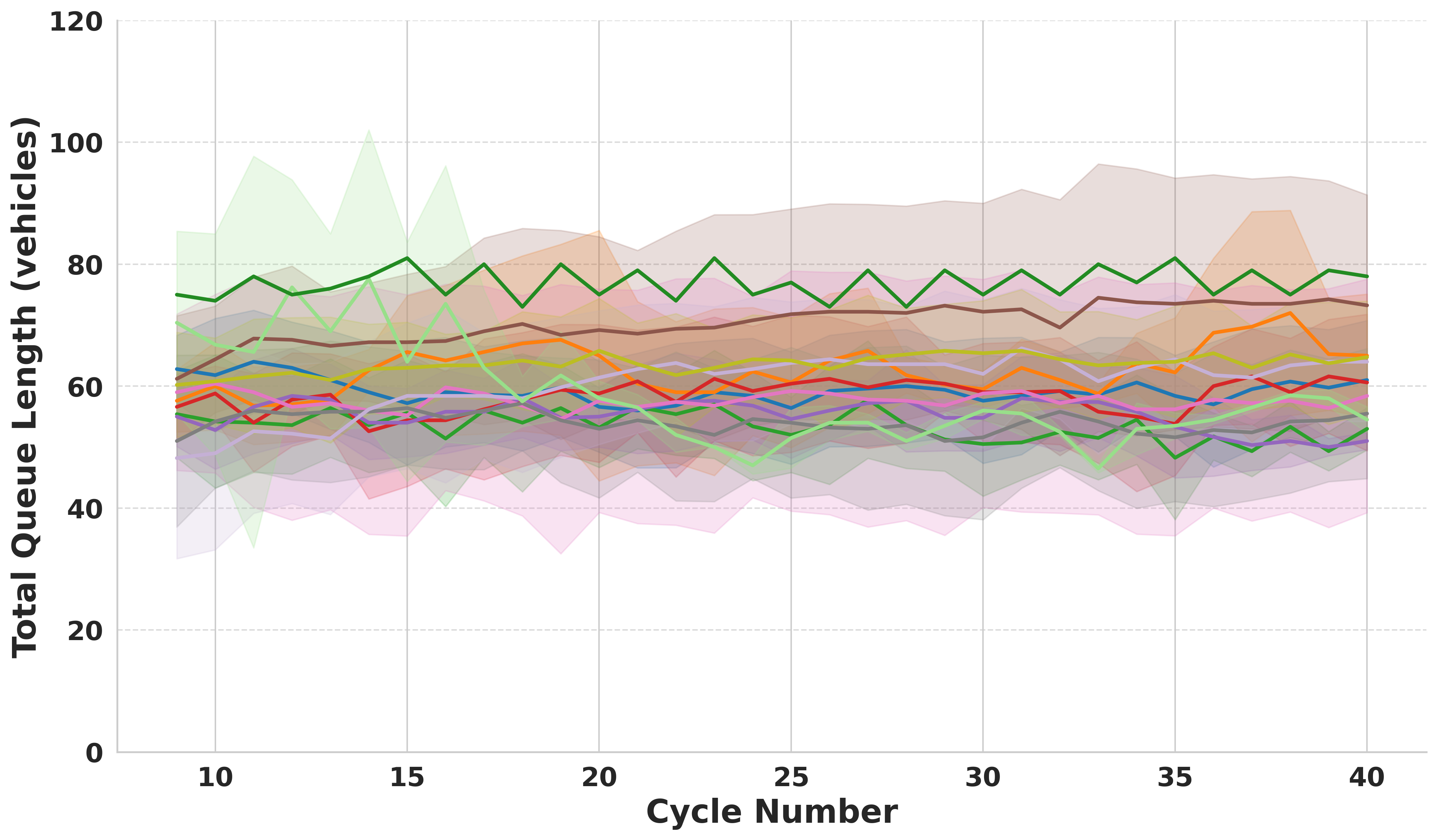}%
\label{fig:total_queue_comparison}
}\hfill
\subfloat[\footnotesize Queue Length vs Phase - High Flow Conditions]{%
\includegraphics[width=0.48\linewidth]{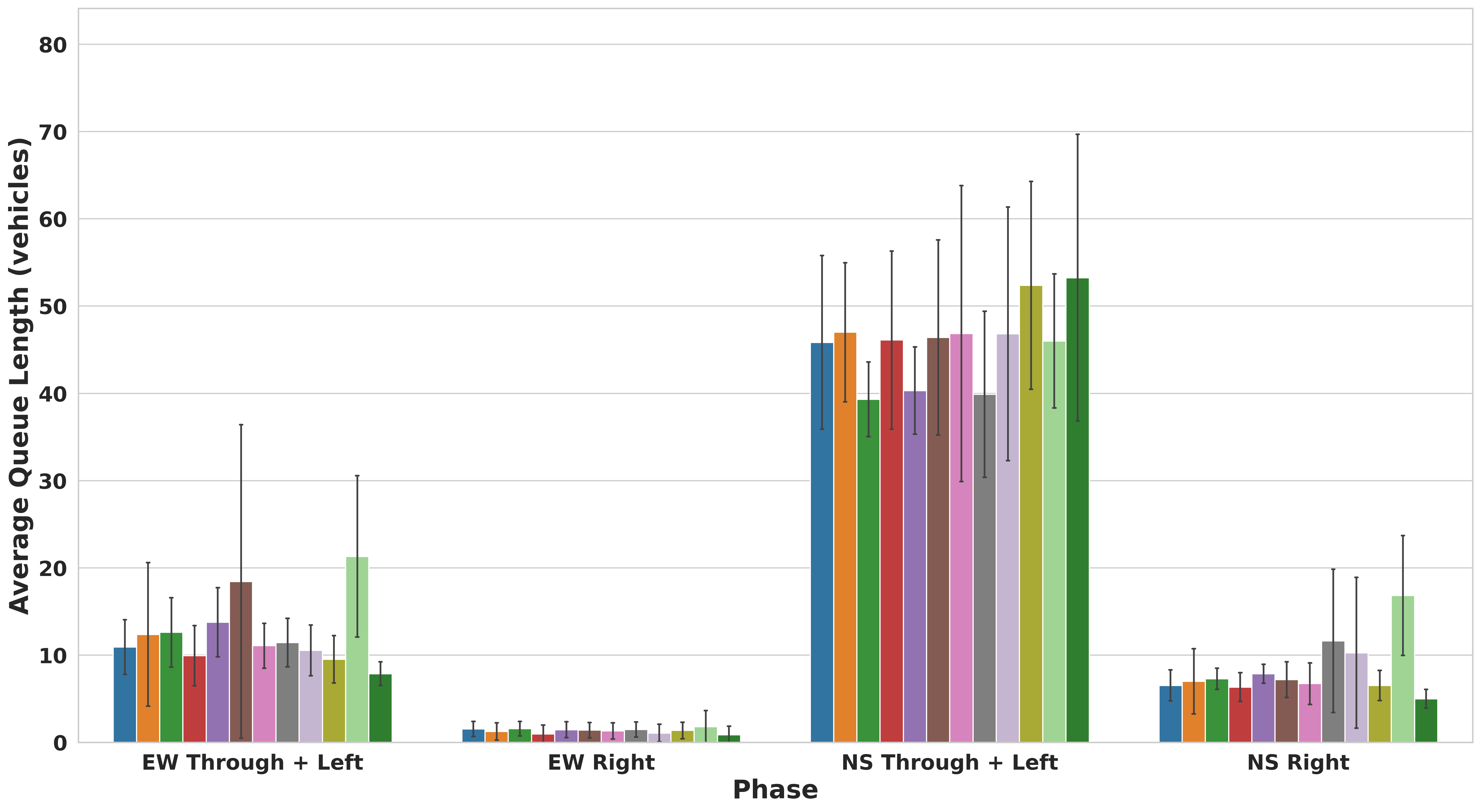}%
\label{fig:queue_length_comparison_high_flow}
}
\vspace{4mm}
\subfloat[\footnotesize Queue Length vs Phase - Medium Flow]{%
\includegraphics[width=0.48\linewidth]{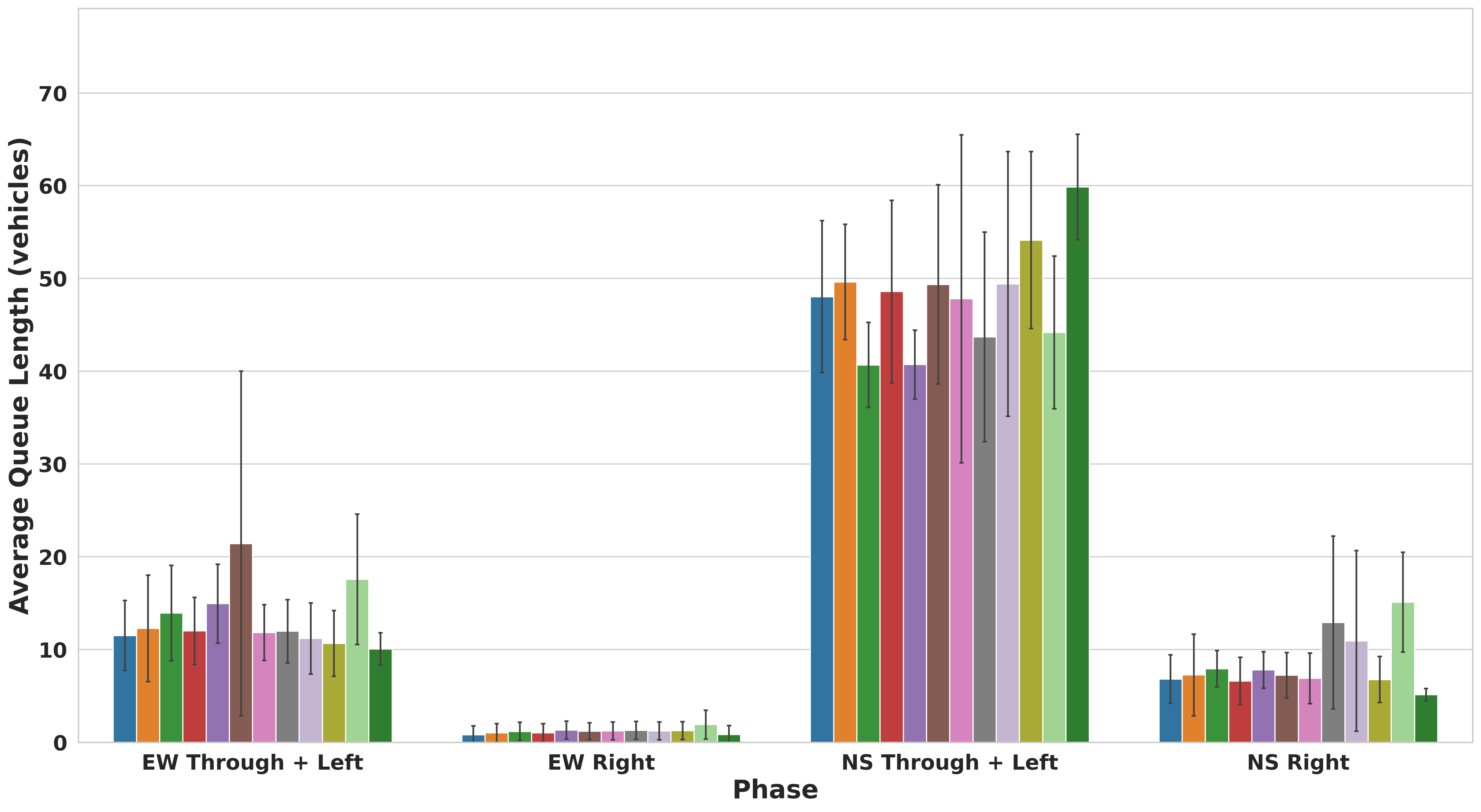}%
\label{fig:queue_length_comparison_medium_flow}
}\hfill
\subfloat[\footnotesize Queue Length vs Phase - Low Flow]{%
\includegraphics[width=0.48\linewidth]{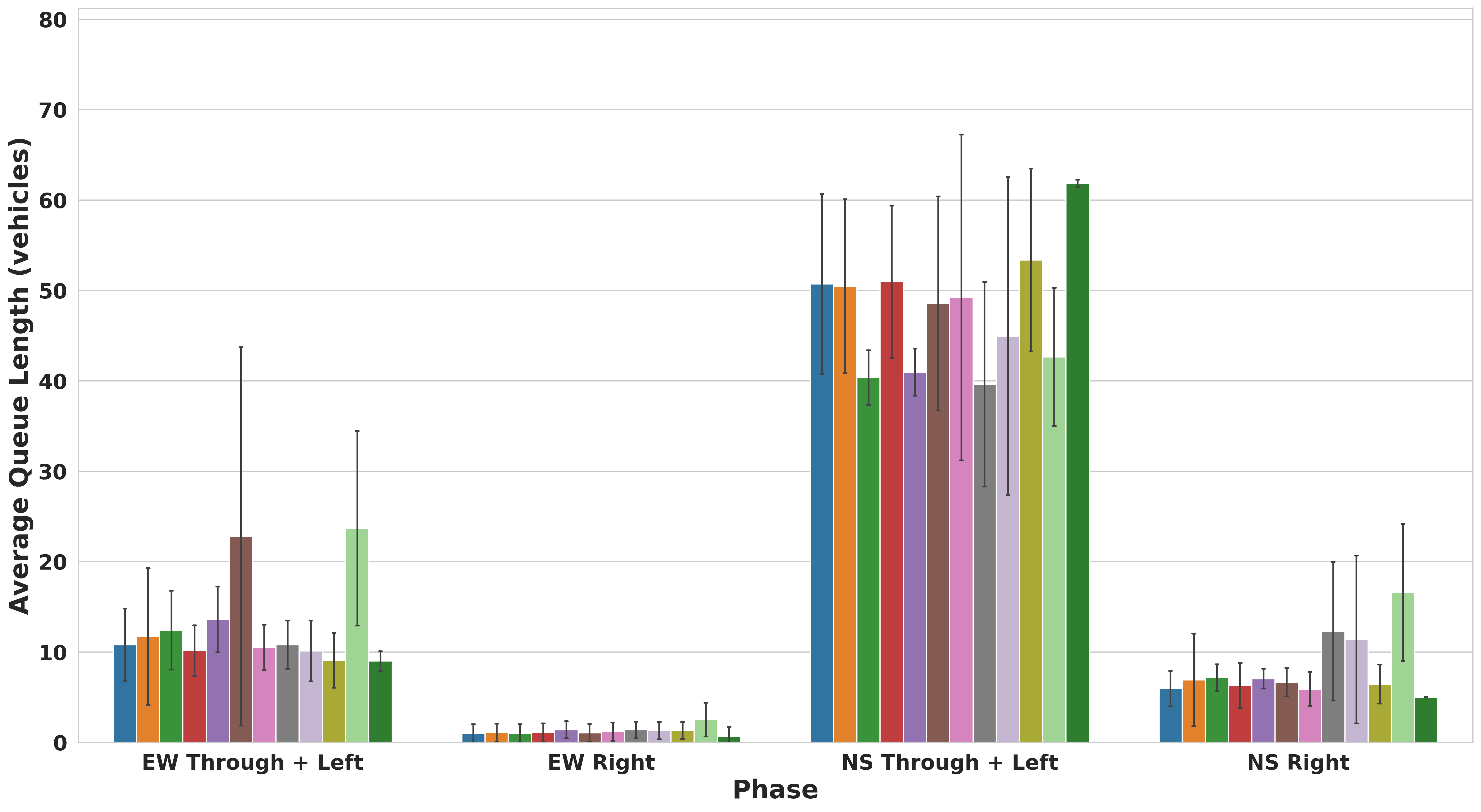}%
\label{fig:low_flow_comparison}
}
\vspace{6mm}
\includegraphics[width=1\linewidth]{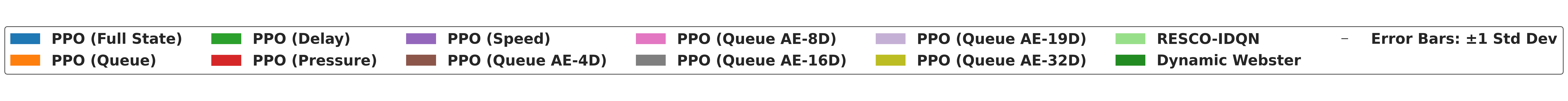}
\caption{Performance evaluation of a PPO controller, configured with various reward functions and state representations, benchmarked against the Dynamic Webster algorithm and RESCO-DQN \cite{NEURIPSDATASETSANDBENCHMARKS2021_f0935e4c} on the LBHO site. Fig. \ref{fig:total_queue_comparison} illustrates the testing performance of the trained controllers, plotting the mean total queue length for each cycle; the shaded region represents $\pm1$ standard deviation across five independent runs (trained with different random seeds). Summarizing the aggregate results, Fig. \ref{fig:queue_length_comparison_high_flow}, \ref{fig:queue_length_comparison_medium_flow}, and \ref{fig:low_flow_comparison} present the phase-wise average queue length as bar charts with error bars ($\pm1$ std. dev.) under high, medium, and low traffic flow conditions, respectively.}
\label{fig:comparison_plots}
\end{figure*}

\subsubsection{Average Speed Reward}
This reward encourages maximizing the average speed of vehicles in the network:
\begin{equation}
R_{\text{speed}}(t) = \frac{1}{M} \sum_{j=1}^{M} v_j(t),
\end{equation}
where $v_j(t)$ is the speed of vehicle $j$ at time $t$ and $M$ is the total number of vehicles.

\begin{figure*}[t]
\centering
\subfloat[\footnotesize Correlation of Queue Length with Green time for the Phase - East-West through and left movement]{%
\includegraphics[width=0.44\linewidth]{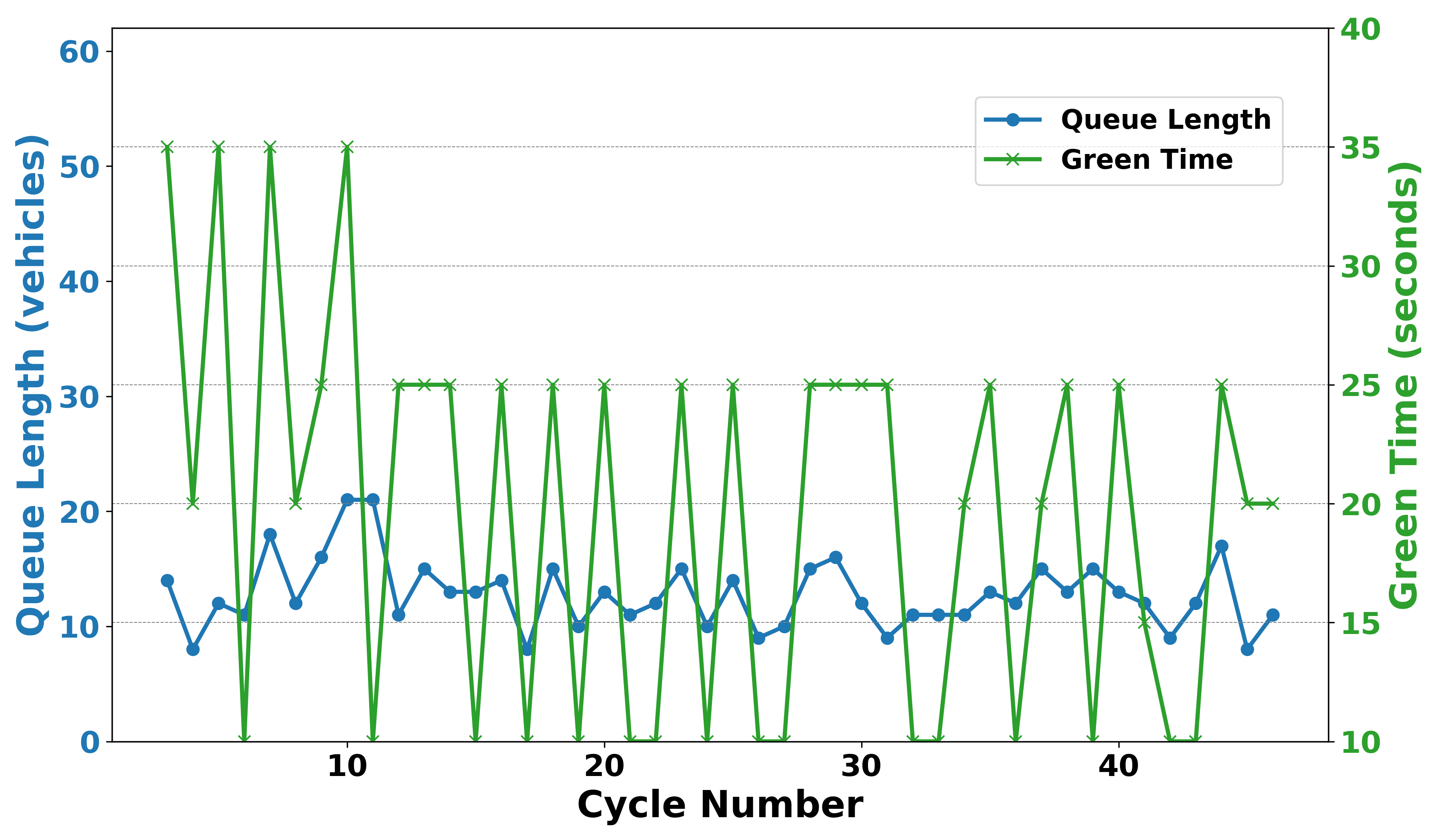}%
\label{fig:east-west_through_left_queue_and_green_vs_cycle_number.png}
}\hfill
\subfloat[\footnotesize Correlation of Queue Length with Green time for the Phase - East-West right movement]{%
\includegraphics[width=0.44\linewidth]{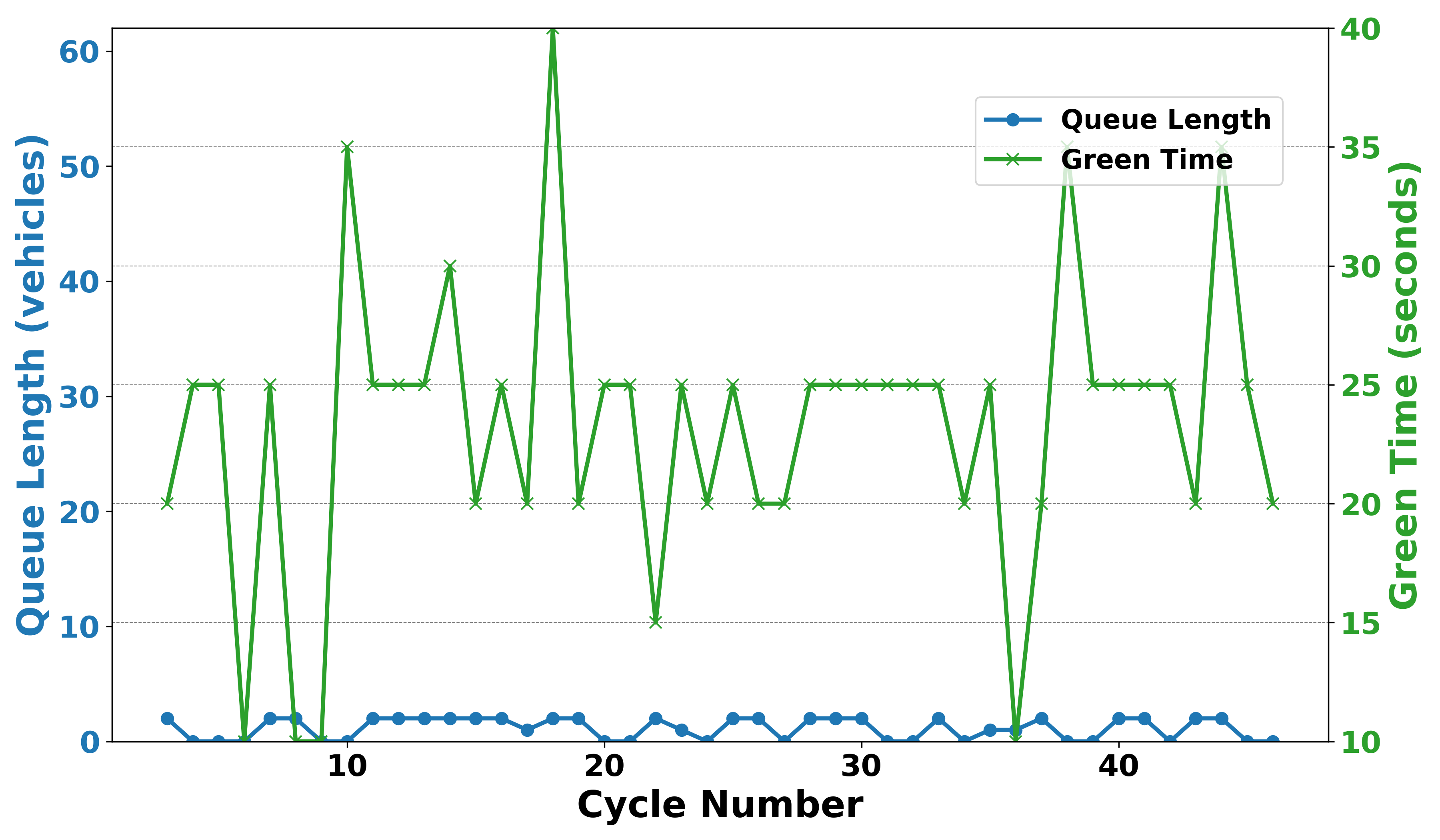}%
\label{fig:east-west_right_queue_and_green_vs_cycle_number.png}
}
\vspace{4mm}
\subfloat[\footnotesize Correlation of Queue Length with Green time for the Phase - North-South through and left movement]{%
\includegraphics[width=0.44\linewidth]{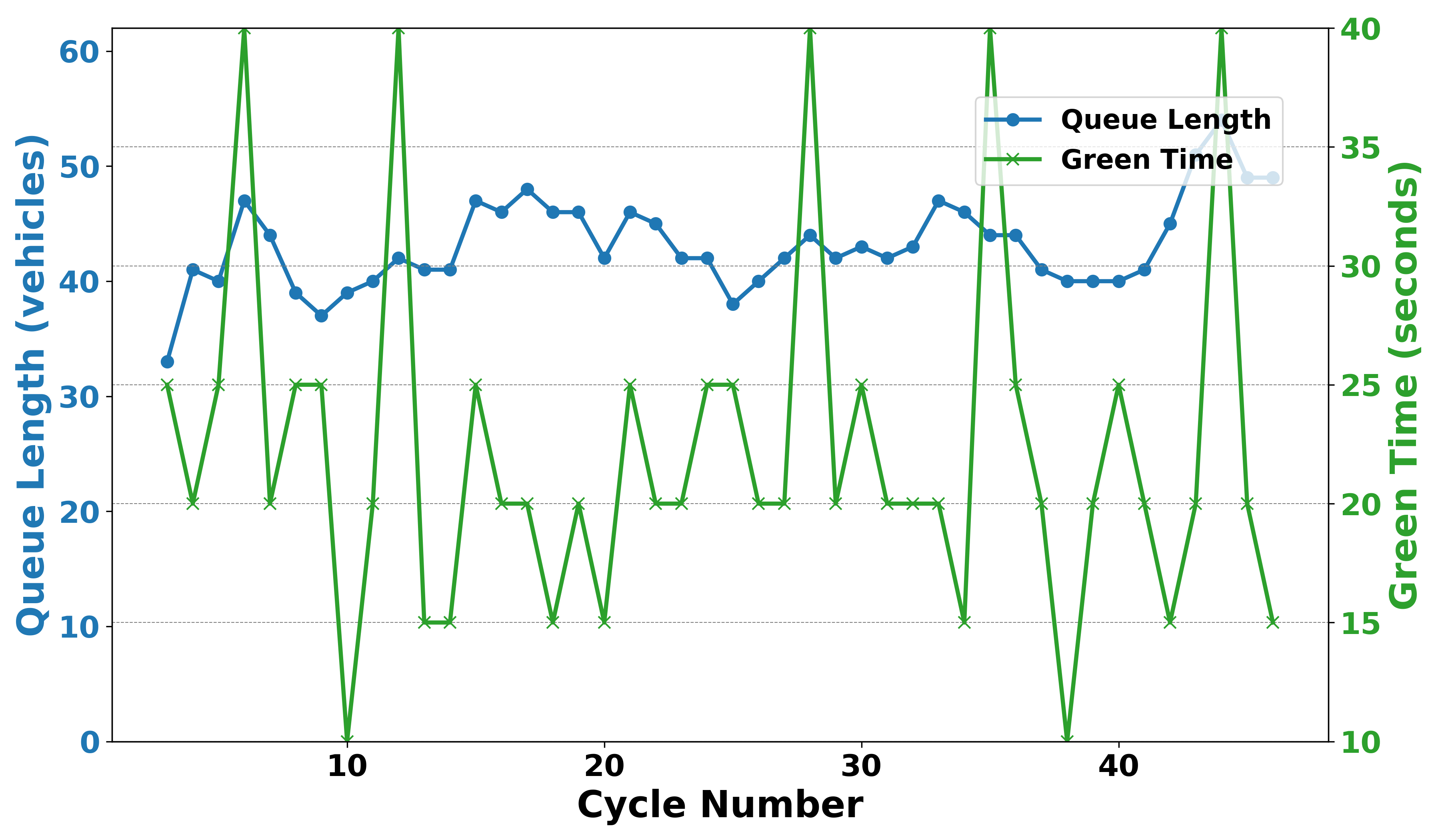}%
\label{fig:north-south_through_left_queue_and_green_vs_cycle_number.png}
}\hfill
\subfloat[\footnotesize Correlation of Queue Length with Green time for the Phase - North-South right movement]{%
\includegraphics[width=0.44\linewidth]{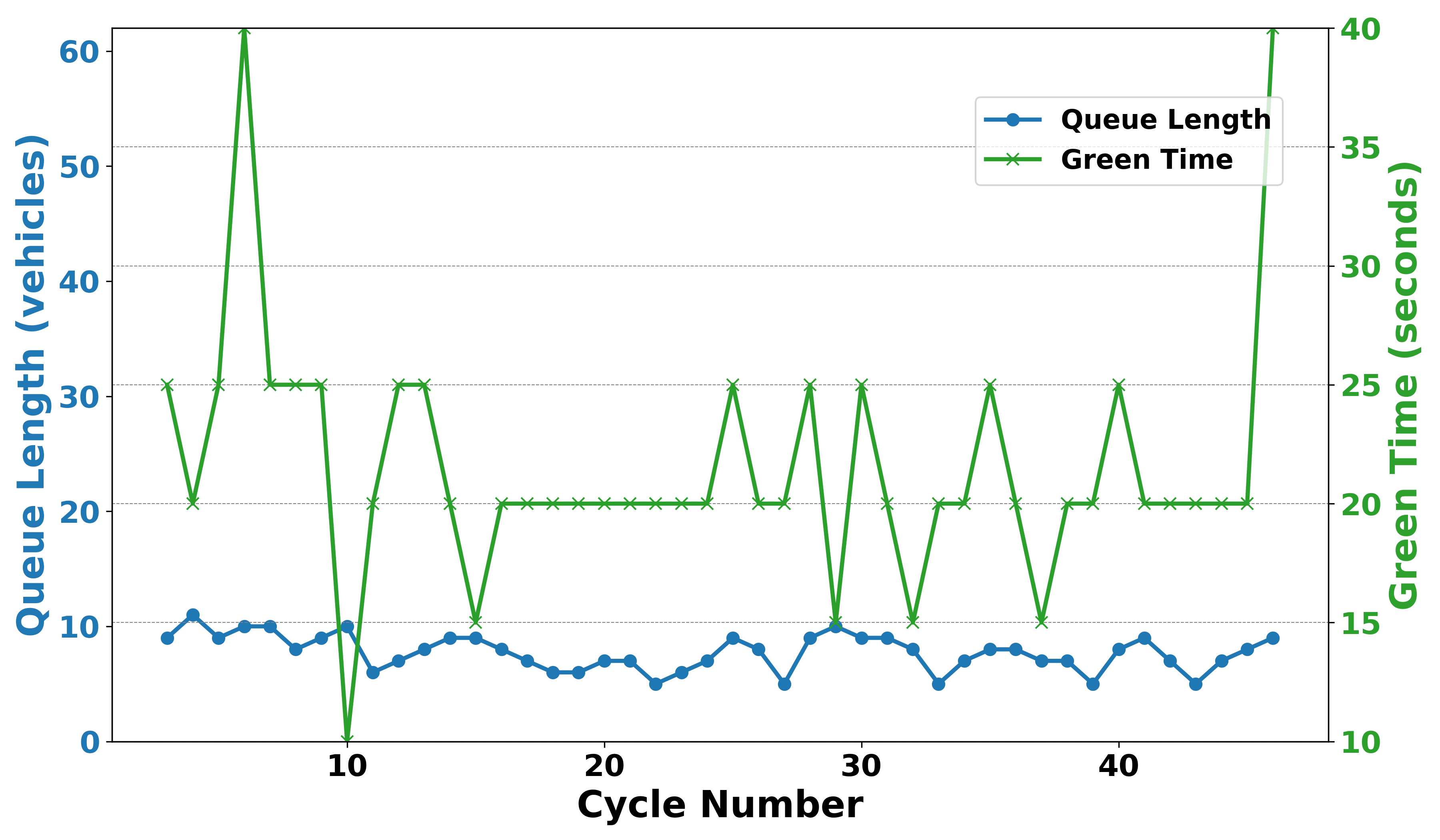}%
\label{fig:north-south_right_queue_and_green_vs_cycle_number.png}
}
\caption{Queue length and green time per phase for the queue based PPO controller on the LBHO site\textcolor{black}{, shown for illustration; the association is not interpreted causally.}}
\label{fig:phase_correlation}
\end{figure*}

\section{Results and Analysis} \label{sec:results}
We describe performance on the LBHO site in Section~\ref{sec:results-homo}, including an autoencoder latent-size sweep under two rewards. Section~\ref{sec:results-hetero} reports performance on the LLHE site, including evaluation under two higher-demand scenarios. Appendix~\ref{app:buildup} reports a cumulative state build-up as a supplementary ablation.

Unless noted otherwise, every configuration is trained with 5 independent runs across different random seeds, and each comparison table marks one reference configuration ($^{\dagger}$) against which the remaining rows are compared (in the build-up tables of Appendix~\ref{app:buildup}, the reference is instead the preceding level). $\Delta$ is the bootstrap 95\% CI on the mean difference from the reference (10{,}000 resamples; seed-paired where the seed sets match, otherwise independent groups), and $g$ is Hedges' effect size. The reference rows are the lowest-mean configuration. With $n=5$ seeds we report intervals and effect sizes rather than significance tests, and no multiple-comparison correction is applied across the roughly 11 comparisons per table; a difference whose interval includes zero is not interpreted as a difference.

\subsection{LBHO Results}
\label{sec:results-homo}
\begin{table}[h!]
\centering
\scriptsize
\caption{Average total queue length per cycle, LBHO site. Reference: PPO (Queue AE-16D)$^{\dagger}$\textcolor{black}{, the lowest-mean configuration. Statistics as defined in Section~\ref{sec:results}}}
\label{tab:queue_length}
\resizebox{\columnwidth}{!}{%
\begin{tabular}{|l|c|c|c|}
\hline
\textbf{Algorithm} & \textbf{Avg. Queue/Cycle} & \textbf{$\Delta$ vs.\ Ref.} & \textbf{Hedges'} \\
 & \textbf{($\mu \pm \sigma$) $\downarrow$} & \textbf{(95\% CI)} & \textbf{$g$} \\
\hline
PPO (Queue AE-16D)$^{\dagger}$ & 53.01 $\pm$ 9.63  & -- & -- \\
PPO (Delay)         & 54.26 $\pm$ 7.59  & [-8.99, +10.16] & +0.13 \\
PPO (Speed)         & 55.36 $\pm$ 5.76  & [-7.07, +10.19] & +0.27 \\
PPO (Queue AE-8D)   & 57.41 $\pm$ 17.37 & [-6.42, +15.23] & +0.28 \\
PPO (Pressure)      & 58.57 $\pm$ 5.75  & [-3.98, +13.60] & +0.63 \\
PPO (Full State)    & 58.76 $\pm$ 7.28  & [-4.21, +14.98] & +0.61 \\
PPO (Queue AE-19D)  & 58.94 $\pm$ 9.21  & [-4.68, +16.55] & +0.57 \\
PPO (Queue AE-32D)  & 62.34 $\pm$ 7.28  & [-4.13, +17.81] & +0.99 \\
PPO (Queue)         & 62.38 $\pm$ 4.73  & [+0.34, +16.70] & +1.12 \\
PPO (Queue AE-4D)   & 68.22 $\pm$ 13.03 & [-2.47, +28.37] & +1.20 \\
RESCO-DQN           & 68.37 $\pm$ 15.75 & [+1.02, +29.59] & +1.06 \\
Dynamic Webster     & 69.01 $\pm$ 0.06  & [+7.55, +21.98] & +2.12 \\
Actuated (SUMO) & 61.52 $\pm$ 0.17 & [-0.08, +17.10] & +1.25 \\
\hline
\end{tabular}%
}

\end{table}

\begin{table}[h!]
\centering
\scriptsize
\caption{Autoencoder reconstruction MSE and resulting queue length by latent dimension, LBHO site, queue based reward. Reference: 16-D$^{\dagger}$\textcolor{black}{ (lowest mean)}}
\label{tab:autoencoder_stats}
\resizebox{\columnwidth}{!}{%
\begin{tabular}{|c|c|c|c|c|c|}
\hline
\textbf{Latent} & \textbf{MSE} & \textbf{MSE} & \textbf{Queue Length} & \textbf{$\Delta$ vs.\ 16-D} & \textbf{Hedges'} \\
\textbf{Dim} & \textbf{Mean $\downarrow$} & \textbf{Std $\downarrow$} & \textbf{(Mean $\pm$ Std) $\downarrow$} & \textbf{(95\% CI)} & \textbf{$g$} \\
\hline
4              & 0.000072 & 0.000011 & 68.22 $\pm$ 13.03 & [-2.47, +28.37] & +1.20 \\
8              & 0.000062 & 0.000002 & 57.41 $\pm$ 17.37 & [-6.42, +15.23] & +0.28 \\
16$^{\dagger}$ & 0.000063 & 0.000009 & 53.01 $\pm$ 9.63  & -- & -- \\
19             & 0.000061 & 0.000004 & 58.94 $\pm$ 9.21  & [-4.68, +16.55] & +0.57 \\
32             & 0.000063 & 0.000009 & 62.34 $\pm$ 7.28  & [-4.13, +17.81] & +0.99 \\
\hline
\end{tabular}%
}
\end{table}

Fig. \ref{fig:comparison_plots} shows the comparison of the performance of the proposed queue based RL algorithms against the benchmarks on the LBHO site. From Fig. \ref{fig:total_queue_comparison} and Table \ref{tab:queue_length}, the queue based algorithm with a 16-D autoencoder state attains the lowest mean total queue length; \textcolor{black}{most pairwise differences among the PPO variants are not statistically distinguishable (95\% CIs include zero), with the exception of PPO (Queue), whose interval just excludes zero, so this ranking is descriptive. All PPO variants have a lower mean queue than Dynamic Webster, PPO (Queue AE-4D) only marginally (68.22 vs.\ 69.01) and comparable to RESCO-DQN (68.37). Relative to Dynamic Webster (69.01), the lowest-mean configuration is about 23\% lower and the unselected PPO (Full State) (58.76) about 15\% lower.} The delay based, average-speed based, pressure based, full-state and other autoencoder based variants exhibit broadly comparable performance. RESCO-DQN has the highest mean queue of the learned baselines. Actuated (SUMO) attains 61.52, comparable to the mid-range PPO variants (its CI relative to the reference includes zero).

\textcolor{black}{\textbf{Reward comparison under matched state:} The Delay, Pressure and Speed variants use the 19-D Full State, so the like-for-like comparison with the queue reward is PPO (Full State) = 58.76. Delay (54.26), Pressure (58.57) and Speed (55.36) all lie within one standard deviation of it, so no difference between the queue based reward and these alternatives is detected at $n=5$. Failing to detect a difference is not evidence of equivalence. The queue reward needs only approach-level queue counts, whereas the delay and pressure rewards need per-vehicle delay or inflow/outflow counts (Section~\ref{subsec:alt-rewards}).}

An analysis of the performance of various autoencoder representations is presented in Table \ref{tab:autoencoder_stats}. A 16-D latent-space autoencoder has the lowest mean total queue length, but its differences to the 8-D, 19-D and 32-D variants are not statistically distinguishable (all 95\% CIs include zero; Table~\ref{tab:autoencoder_stats}). \textcolor{black}{Reconstruction MSE is small and similar across latent sizes.}

An analysis of the queue based PPO-RL controller's performance is presented in Fig. \ref{fig:phase_correlation}, which \textcolor{black}{illustrates} the co-movement between the green time allocated and the corresponding queue length for each phase (we report this qualitatively; the direction of causality is not identified here, since longer green times also shape future queues). Plots show that the green time of a phase tends to track the queue observed. A higher queue in a phase is typically associated with more green time given to the phase. This could be seen in detail for the critical phases north-south through and left movements, Fig. \ref{fig:north-south_through_left_queue_and_green_vs_cycle_number.png}, and east-west through and left movements, Fig. \ref{fig:east-west_through_left_queue_and_green_vs_cycle_number.png}, where an increase in queue length coincides with an increase in the green time allocated.

Fig. \ref{fig:queue_length_comparison_high_flow}, Fig. \ref{fig:queue_length_comparison_medium_flow}, and Fig. \ref{fig:low_flow_comparison} present a comparison of the proposed algorithm with benchmark approaches under high, medium, and low traffic flow conditions, respectively. The north-south through and left movement phase is identified as the most critical, experiencing the highest volumes, followed by the east-west through and left movement phase. The plots show that the queue based RL approach with 16-D autoencoder representation attains the lowest or near-lowest queue in the critical phases. Although the dynamic Webster design results in reduced queue lengths for the east-west through and left movement phase, it causes increased queue lengths in the north-south through and left phase, contributing to its overall sub-optimal aggregate performance. A similar pattern is observed for RESCO-DQN, where improvement in one phase (north-south through and left) is offset by degradation in another (east-west through and left), reducing overall effectiveness.

\subsubsection{Representation Sweep Under an Alternative Reward}
\label{sec:results-homo-delay}
The autoencoder latent-size sweep above is trained exclusively under the queue based reward (Eq.~\eqref{eq:reward}). Table~\ref{tab:autoencoder_stats_delay} repeats the same sweep under the delay based reward of Section~\ref{sec:baselines}.

\begin{table}[h!]
\centering
\scriptsize
\caption{As Table~\ref{tab:autoencoder_stats}, under the \textbf{delay based reward}. Reference: 19-D$^{\dagger}$ (lowest mean under this reward)}
\label{tab:autoencoder_stats_delay}
\resizebox{\columnwidth}{!}{%
\begin{tabular}{|c|c|c|c|c|c|}
\hline
\textbf{Latent} & \textbf{MSE} & \textbf{MSE} & \textbf{Queue Length} & \textbf{$\Delta$ vs.\ Ref.} & \textbf{Hedges'} \\
\textbf{Dim} & \textbf{Mean $\downarrow$} & \textbf{Std $\downarrow$} & \textbf{(Mean $\pm$ Std) $\downarrow$} & \textbf{(95\% CI)} & \textbf{$g$} \\
\hline
4              & $6.4\times10^{-5}$ & $7.3\times10^{-6}$ & 56.27 $\pm$ 10.85 & [-1.38, +9.47]   & +0.34 \\
8              & $6.1\times10^{-5}$ & $4.2\times10^{-6}$ & 53.85 $\pm$ 15.64 & [-12.96, +14.59] & +0.06 \\
16             & $6.1\times10^{-5}$ & $9.0\times10^{-6}$ & 62.80 $\pm$ 8.61  & [+4.61, +16.52]  & +1.22 \\
19$^{\dagger}$ & $5.8\times10^{-5}$ & $4.7\times10^{-6}$ & 53.03 $\pm$ 5.52  & -- & -- \\
32             & $5.9\times10^{-5}$ & $4.9\times10^{-6}$ & 59.65 $\pm$ 9.53  & [-2.90, +15.60]  & +0.77 \\
\hline
\end{tabular}%
}
\end{table}

Under the delay based reward, the lowest mean queue is obtained at 19-D (53.03), with 8-D essentially tied (53.85); only 16-D is clearly worse (95\% CI $[+4.6, +16.5]$). \textcolor{black}{The lowest-mean latent size therefore differs from that under the queue based reward (16-D). Given the interval widths, we do not infer a systematic dependence of performance on latent size.}

\subsection{LLHE Results}
\label{sec:results-hetero}
On the LLHE site, all trained RL configurations have a substantially lower mean queue than the Dynamic Webster baseline under nominal test flows (Table~\ref{tab:queue_length_hetero_nominal}) \textcolor{black}{; the baseline caveat of Section~\ref{sec:baselines} applies to the size of this gap.}

\begin{table}[h!]
\centering
\scriptsize
\setlength{\tabcolsep}{3pt}
\caption{Average total queue length per cycle under nominal test flows, LLHE site. Reference: PPO (Queue AE-8D)$^{\dagger}$\textcolor{black}{, the lowest-mean configuration (chosen after the fact)}}
\label{tab:queue_length_hetero_nominal}
\resizebox{\columnwidth}{!}{%
\begin{tabular}{|l|c|c|c|c|c|}
\hline
\textbf{Controller} & \textbf{Avg Queue} & \textbf{Wait} & \textbf{Speed} & \textbf{$\Delta$ vs.\ Ref.} & \textbf{Hedges'} \\
 & \textbf{(Mean $\pm$ Std) $\downarrow$} & \textbf{(s) $\downarrow$} & \textbf{(m/s) $\uparrow$} & \textbf{(95\% CI)} & \textbf{$g$} \\
\hline
PPO (Queue AE-16D) & 30.67 $\pm$ 5.46 & 0.80 & 9.53 & [-2.03, +6.02] & +0.45 \\
PPO (Delay)        & 34.85 $\pm$ 2.89 & 0.83 & 9.52 & [+3.21, +9.15] & +2.47 \\
\textbf{PPO (Queue AE-8D)}$^{\dagger}$ & 28.68 $\pm$ 1.36 & 0.73 & 9.62 & -- & -- \\
PPO (Pressure)     & 35.96 $\pm$ 5.02 & 0.88 & 9.52 & [+3.83, +10.49] & +1.79 \\
PPO (Full State)   & 35.24 $\pm$ 7.22 & 0.87 & 9.39 & [+1.67, +13.35] & +1.14 \\
PPO (Queue AE-19D) & 29.48 $\pm$ 2.67 & 0.74 & 9.60 & [-1.66, +3.28] & +0.34 \\
PPO (Queue AE-32D) & 32.27 $\pm$ 2.20 & 0.85 & 9.43 & [+1.74, +5.33] & +1.78 \\
PPO (Queue)        & 33.18 $\pm$ 4.50 & 0.75 & 9.58 & [+0.76, +7.66] & +1.22 \\
PPO (Queue AE-4D)  & 30.10 $\pm$ 2.45 & 0.72 & 9.60 & [-0.46, +4.18] & +0.65 \\
RESCO-DQN          & 33.27 $\pm$ 4.57 & 0.37 & 9.76 & [+1.82, +7.36] & +1.23 \\
Actuated (SUMO)    & 36.62 $\pm$ 2.39 & 0.95 & 9.46 & [+5.70, +10.00] & +3.69 \\
Dynamic Webster    & 105.60 $\pm$ 4.45 & 5.19 & 8.06 & [+73.05, +80.26] & +21.10 \\
\hline
\end{tabular}%
}
\end{table}

\textcolor{black}{Relative to Dynamic Webster (105.60) in Table~\ref{tab:queue_length_hetero_nominal}, the RL configurations have 66--73\% lower mean queue. Among the RL variants, the intervals against the reference (AE-8D, itself selected as the lowest mean) exclude zero for the raw-queue state, the Full State, AE-32D, the Delay and Pressure variants (which use the Full State), RESCO-DQN, and Actuated (SUMO), and include zero for AE-4D, AE-16D and AE-19D. Under the matched Full State, the Delay (34.85), Pressure (35.96) and Queue (35.24) rewards are within one standard deviation of each other, so no difference between the rewards is detected at $n=5$.}

\subsubsection{\textcolor{black}{Evaluation Under Demand Surges}}
\label{sec:surge}
All trained controllers (including Dynamic Webster) are re-evaluated, without further training, on two higher-demand route files generated from the nominal test route:
\begin{itemize}
\item \textbf{Surge-1 (Uniform):} every flow scaled by a constant factor of +35\%.
\item \textbf{Surge-2 (Random, per-flow):} every flow independently scaled by a factor sampled uniformly from +10\% to +60\% (seed-fixed for reproducibility), producing an unbalanced demand pattern across approaches.
\end{itemize}

\paragraph{Surge-1: Uniform +35\% Demand}
Table \ref{tab:surge1_summary} compares the performance of several controllers under the uniform Surge-1 demand condition using three evaluation metrics: average queue length, average waiting time, and average speed.

\begin{table}[h!]
\centering
\scriptsize
\setlength{\tabcolsep}{3pt}
\caption{Controller performance under Surge-1 (uniform +35\%), LLHE site. Reference: PPO (Queue AE-8D)$^{\dagger}$; statistics on average queue }
\label{tab:surge1_summary}
\resizebox{\columnwidth}{!}{%
\begin{tabular}{|l|c|c|c|c|c|}
\hline
\textbf{Controller} & \textbf{Avg Queue} & \textbf{Wait} & \textbf{Speed} & \textbf{$\Delta$ vs.\ Ref.} & \textbf{Hedges'} \\
 & \textbf{(Mean $\pm$ Std) $\downarrow$} & \textbf{(s) $\downarrow$} & \textbf{(m/s) $\uparrow$} & \textbf{(95\% CI)} & \textbf{$g$} \\
\hline
PPO (Queue AE-16D) & 29.77 $\pm$ 5.00 & 0.73 & 9.60 & [-2.05, +5.29] & +0.40 \\
PPO (Delay)        & 35.68 $\pm$ 4.64 & 0.82 & 9.53 & [+3.31, +11.76] & +1.99 \\
\textbf{PPO (Queue AE-8D)}$^{\dagger}$ & \textbf{28.15 $\pm$ 1.31} & 0.67 & \underline{9.68} & -- & -- \\
PPO (Pressure)     & 36.18 $\pm$ 4.43 & 0.83 & 9.58 & [+5.10, +10.98] & +2.22 \\
PPO (Full State)   & 40.21 $\pm$ 14.45 & 0.70 & 9.43 & [+2.52, +24.79] & +1.06 \\
PPO (Queue AE-19D) & \underline{28.96 $\pm$ 2.68} & 0.68 & 9.66 & [-1.78, +3.40] & +0.35 \\
PPO (Queue AE-32D) & 31.46 $\pm$ 3.05 & 0.79 & 9.52 & [+0.96, +5.59] & +1.27 \\
PPO (Queue)        & 33.03 $\pm$ 5.42 & 0.71 & 9.63 & [+0.49, +8.87] & +1.12 \\
PPO (Queue AE-4D)  & 29.64 $\pm$ 2.35 & 0.66 & 9.65 & [-0.55, +4.07] & +0.71 \\
RESCO-DQN          & 34.94 $\pm$ 4.86 & \textbf{0.38} & \textbf{9.83} & [+3.93, +9.89] & +1.72 \\
Actuated (SUMO)    & 40.08 $\pm$ 1.97 & 1.02 & 9.51 & [+9.96, +13.70] & +6.44 \\
Dynamic Webster    & 108.84 $\pm$ 1.88 & 6.74 & 7.84 & [+78.80, +82.37] & +45.02 \\
\hline
\end{tabular}%
}
\end{table}

\begin{figure*}[t]
\centering
\subfloat[\footnotesize Total Queue Length vs Cycle Number]{%
\includegraphics[width=0.48\linewidth]{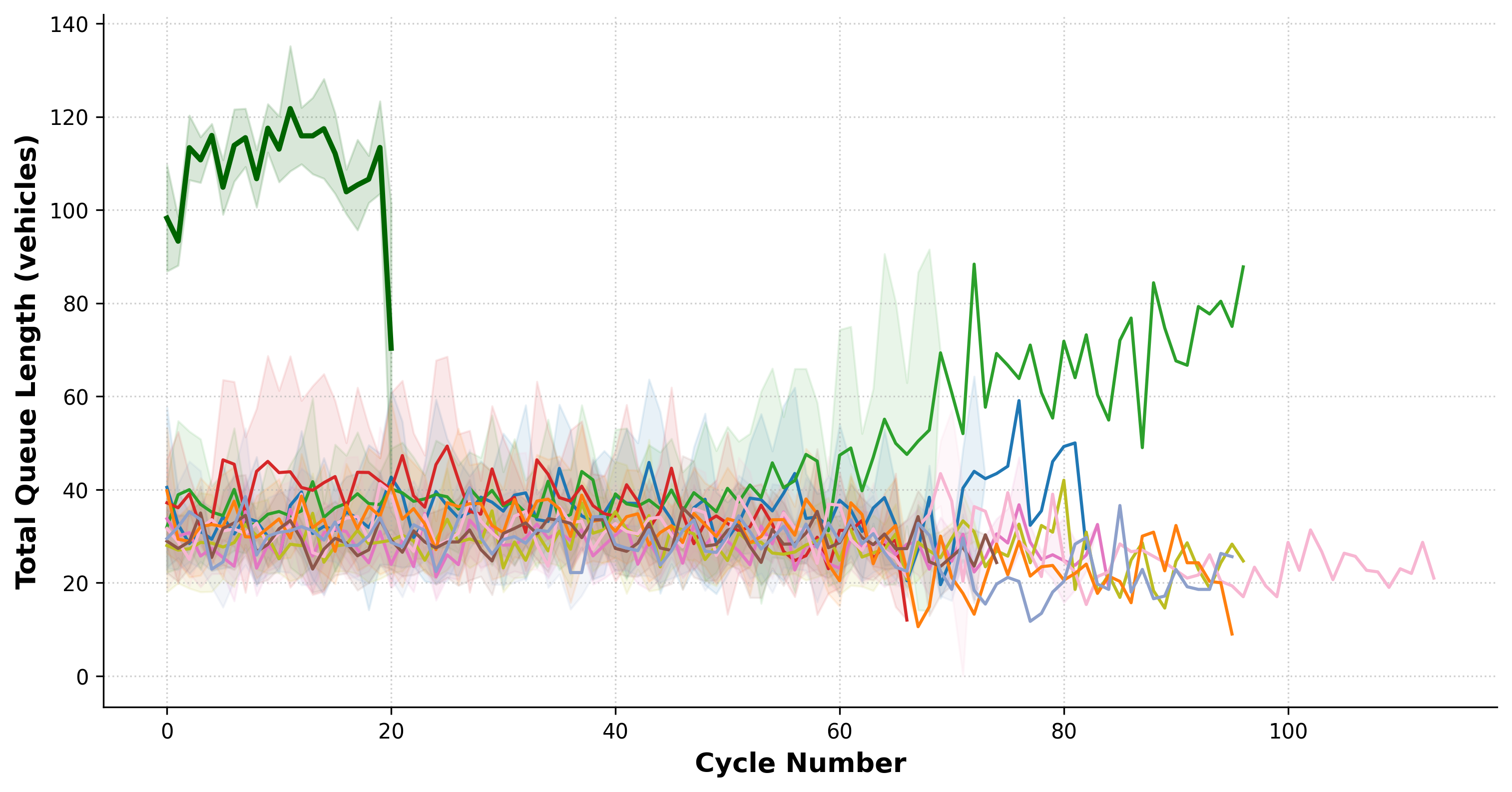}%
\label{fig:total_queue_comparison_surge1}
}\hfill
\subfloat[\footnotesize Queue Length vs Phase, High Flow]{%
\includegraphics[width=0.48\linewidth]{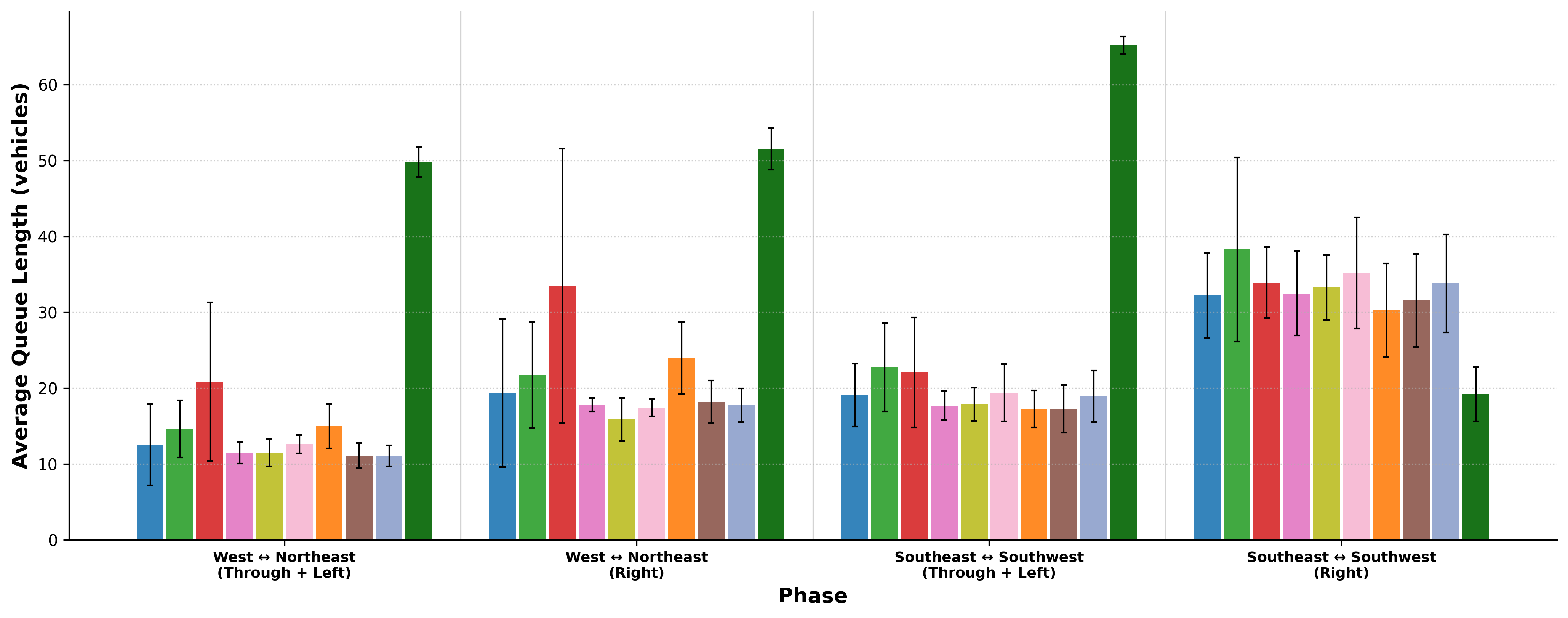}%
\label{fig:queue_length_comparison_high_flow_surge1}
}
\vspace{4mm}
\subfloat[\footnotesize Queue Length vs Phase, Medium Flow]{%
\includegraphics[width=0.48\linewidth]{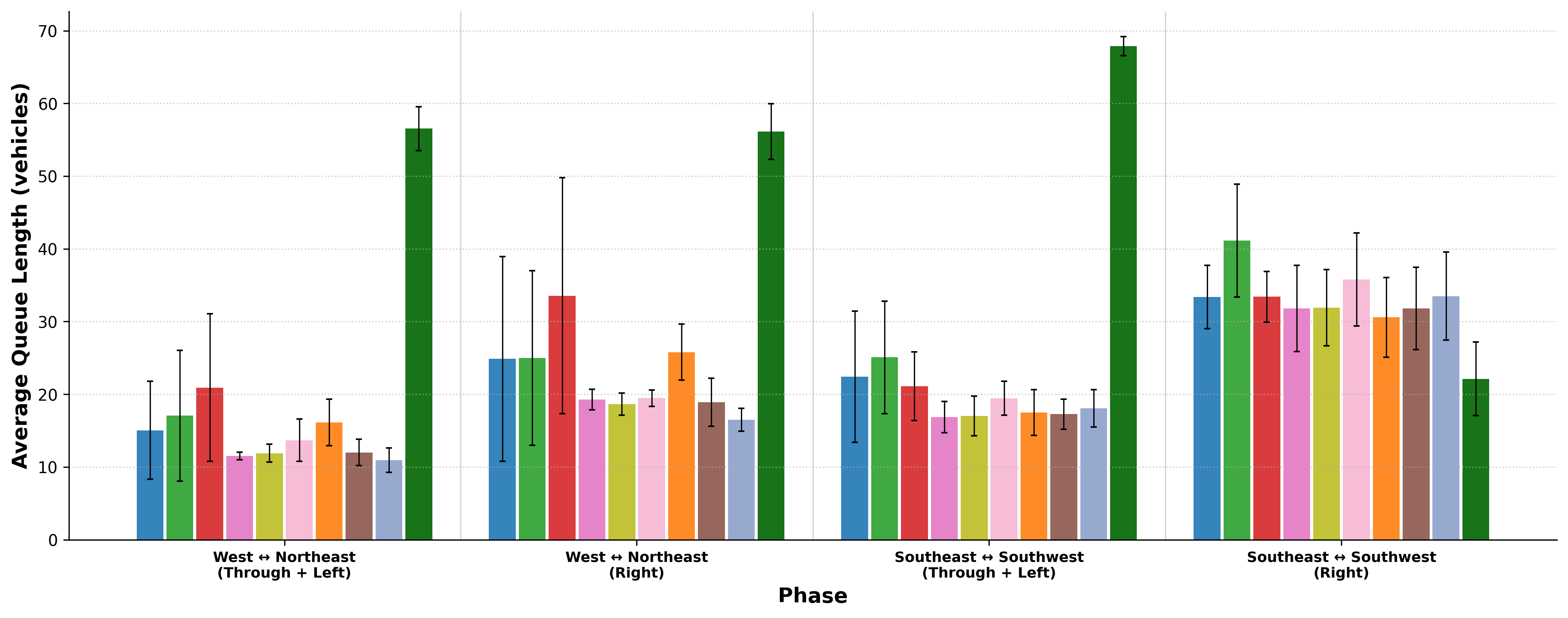}%
\label{fig:queue_length_comparison_medium_flow_surge1}
}\hfill
\subfloat[\footnotesize Queue Length vs Phase, Low Flow]{%
\includegraphics[width=0.48\linewidth]{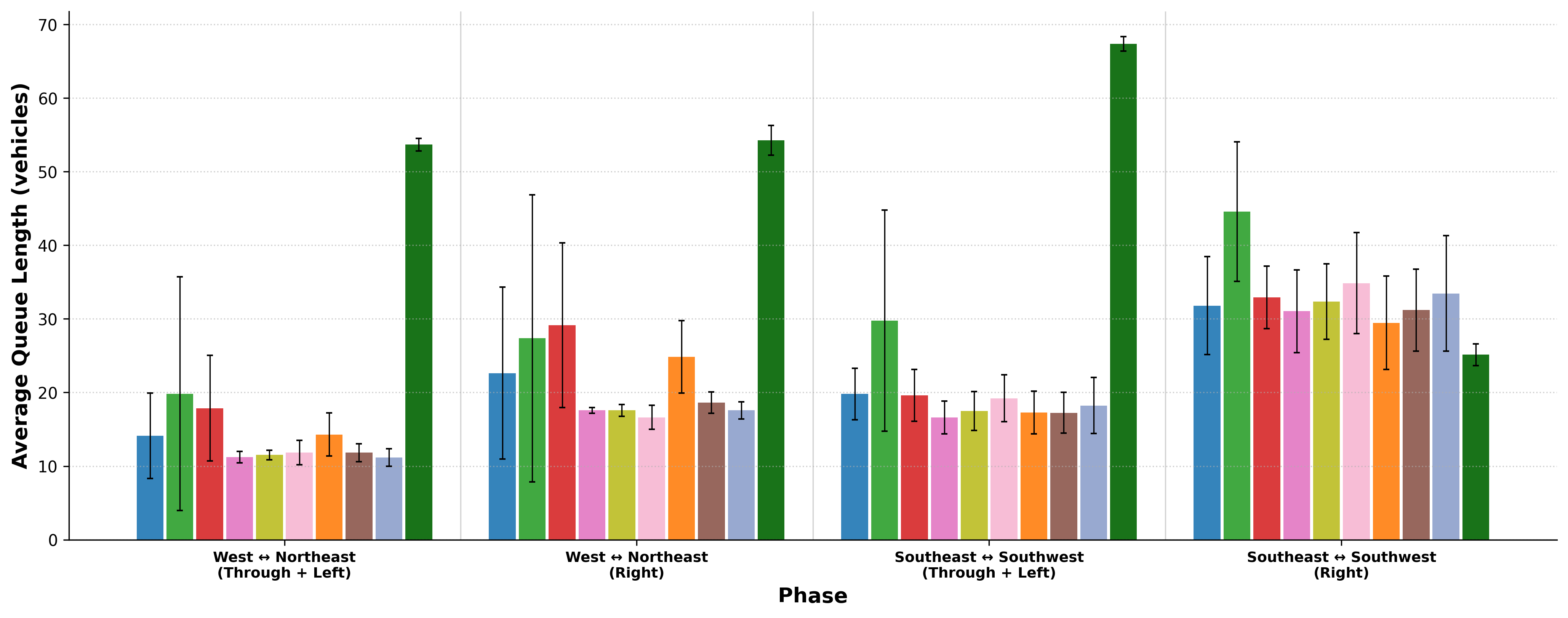}%
\label{fig:low_flow_comparison_surge1}
}
\vspace{6mm}
\includegraphics[width=\linewidth]{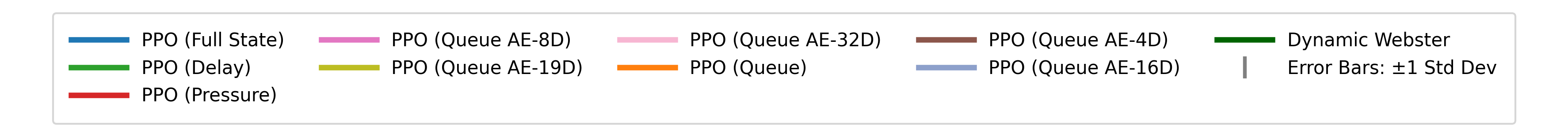}
\caption{Phase-wise queue length evaluation on the LLHE site under Surge-1 (uniform +35\%). (a) Mean total queue length per cycle ($\pm1$ std.\ dev.\ across five seeds). (b)--(d) Per-phase average queue lengths under high, medium, and low flow conditions. Overall, RL controllers outperform Dynamic Webster across the different phases.}
\label{fig:comparison_plots_surge1}
\end{figure*}

\begin{figure*}[t]
\centering
\subfloat[\footnotesize Total Queue Length vs Cycle Number]{%
\includegraphics[width=0.48\linewidth]{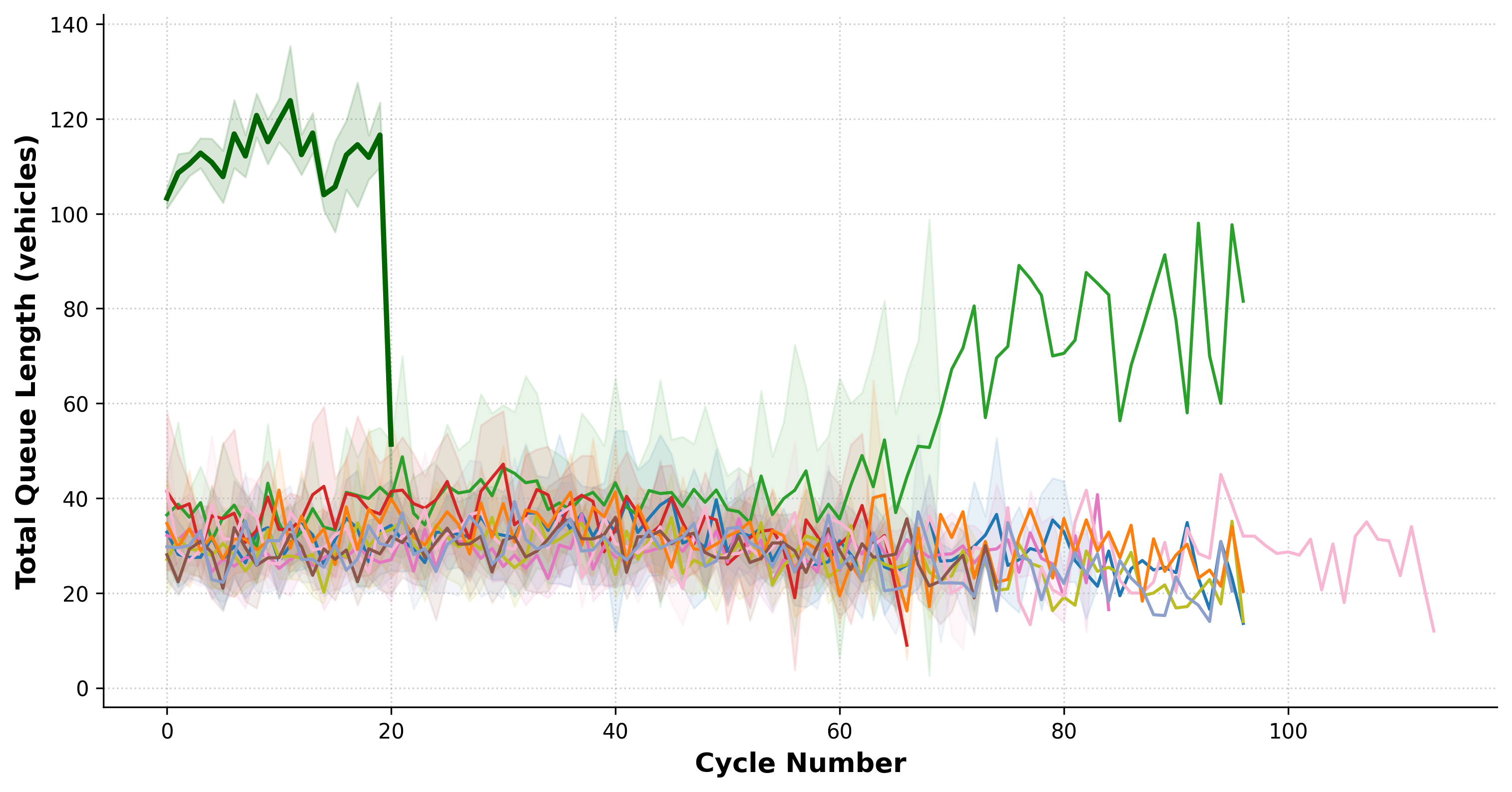}%
\label{fig:total_queue_comparison_surge2}
}\hfill
\subfloat[\footnotesize Queue Length vs Phase, High Flow]{%
\includegraphics[width=0.48\linewidth]{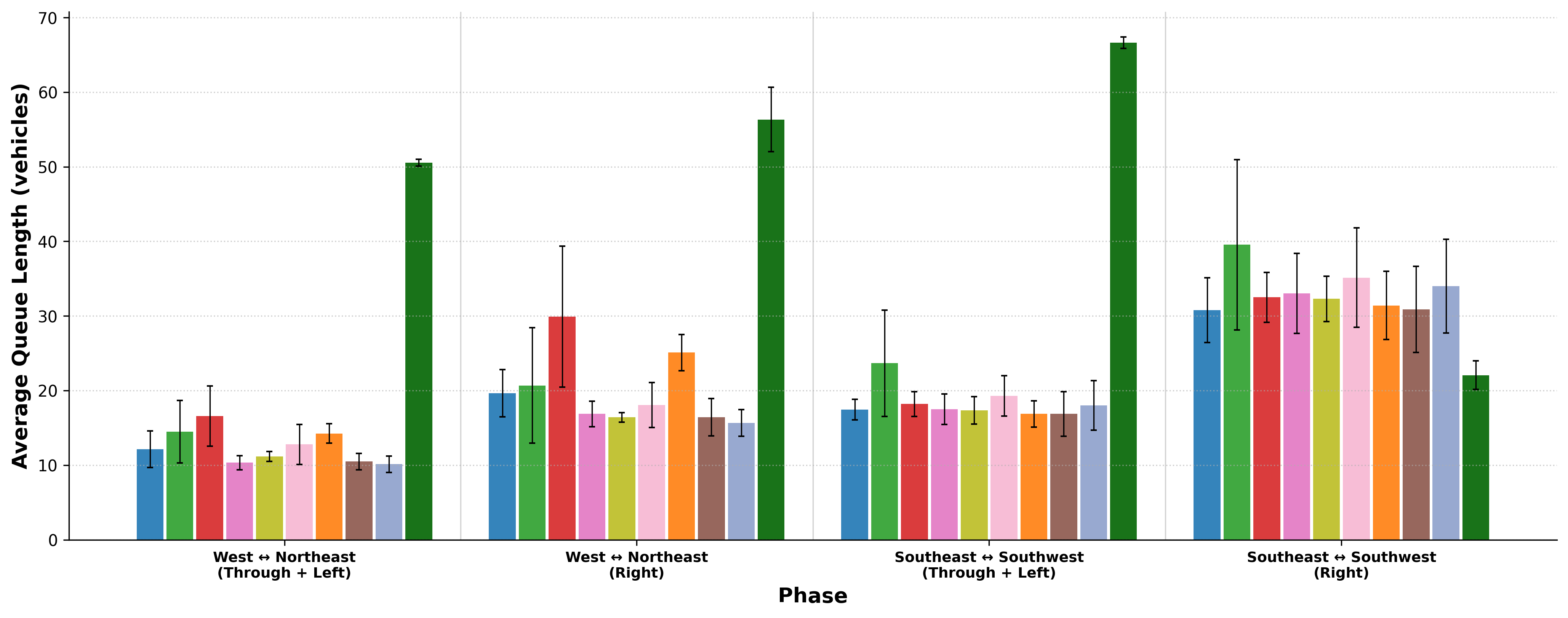}%
\label{fig:queue_length_comparison_high_flow_surge2}
}
\vspace{4mm}
\subfloat[\footnotesize Queue Length vs Phase, Medium Flow]{%
\includegraphics[width=0.48\linewidth]{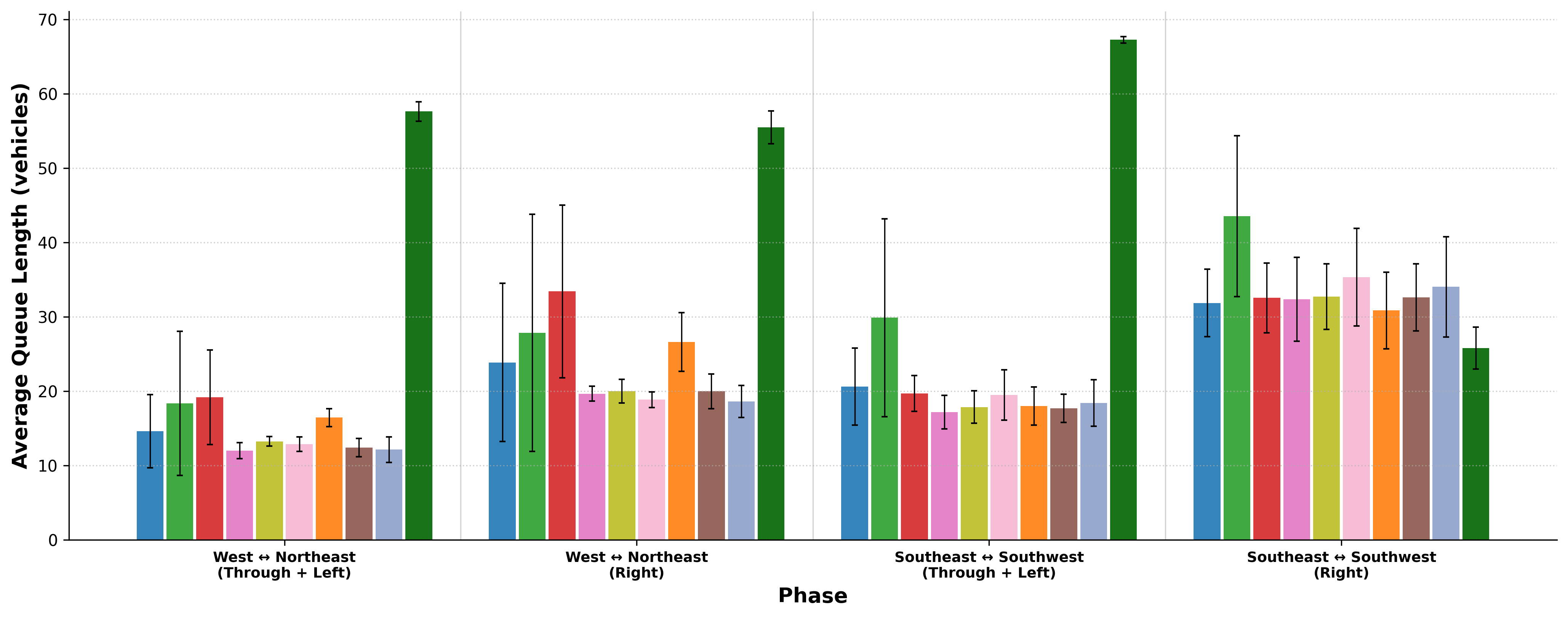}%
\label{fig:queue_length_comparison_medium_flow_surge2}
}\hfill
\subfloat[\footnotesize Queue Length vs Phase, Low Flow]{%
\includegraphics[width=0.48\linewidth]{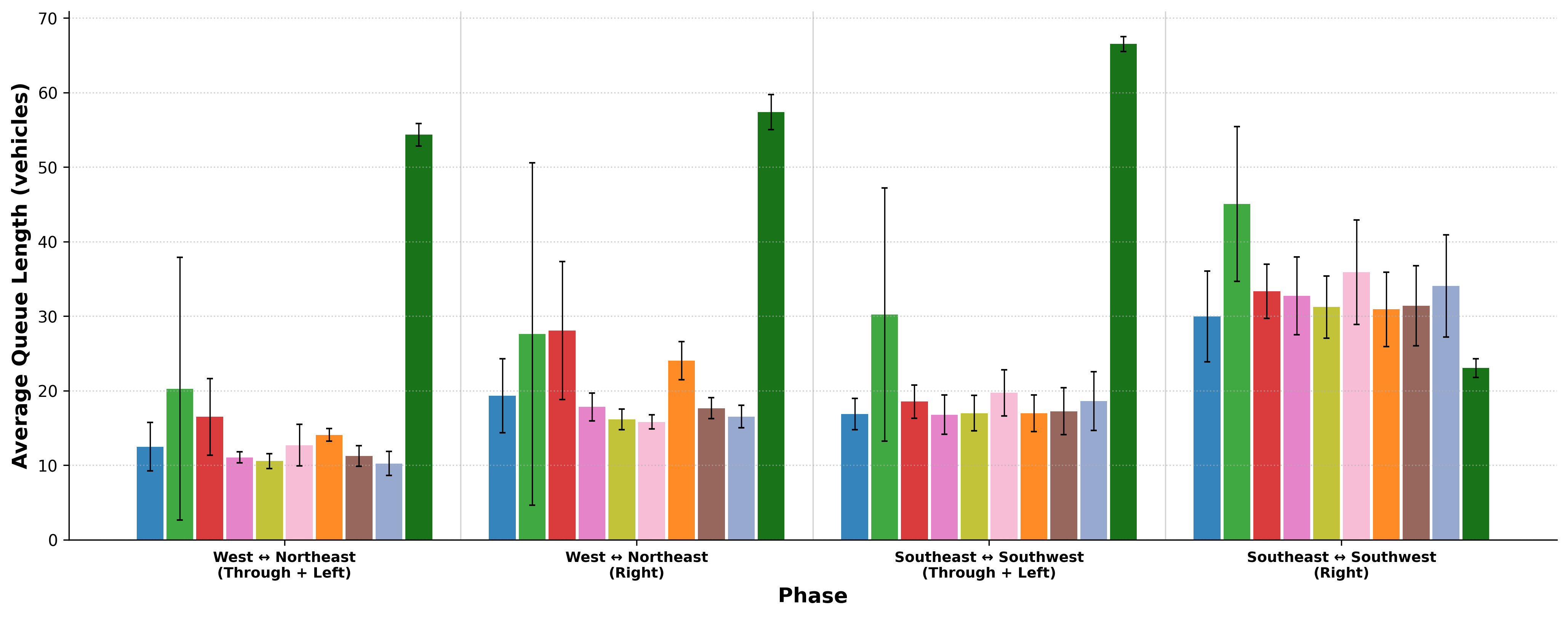}%
\label{fig:low_flow_comparison_surge2}
}
\vspace{6mm}
\includegraphics[width=\linewidth]{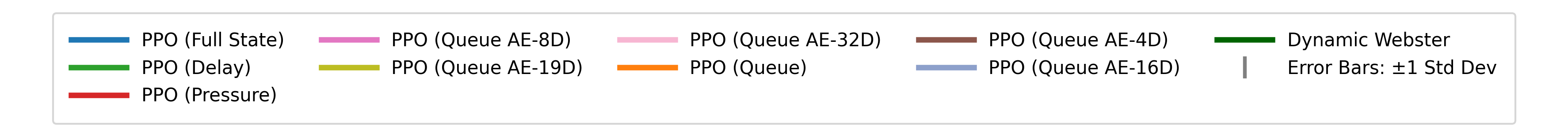}
\caption{Phase-wise queue length evaluation on the LLHE site under Surge-2 (random per-flow +10\%--60\%). Layout matches Fig.~\ref{fig:comparison_plots_surge1}. The ranking across controllers is consistent with Surge-1.}
\label{fig:comparison_plots_surge2}
\end{figure*}

Under a uniform +35\% surge, every RL configuration has a lower mean queue than the Dynamic Webster baseline (63\%--74\% lower, computed from the entries of Table \ref{tab:surge1_summary}). \textcolor{black}{However, under Surge-1, the mean queue remains within one standard deviation of its nominal-flow value for every controller except actuated control, including Dynamic Webster (105.60 $\to$ 108.84), and is slightly lower than nominal for several RL configurations (e.g.\ AE-8D: 28.68 $\to$ 28.15). Queues thus change little under the surge for any controller. We therefore report these runs only as showing that the relative standing of the controllers is unchanged, and we draw no conclusion about generalization or robustness; the surge may simply not alter the quantity measured by the per-cycle queue metric.} Figs. \ref{fig:comparison_plots_surge1} and \ref{fig:comparison_plots_surge2} (Appendix) show the phase-wise breakdowns, with Webster producing larger queue lengths than the RL controllers throughout the simulation.

\paragraph{Surge-2: Random Per-Flow +10\%-60\% Demand}
Table \ref{tab:surge2_summary} compares the performance of several controllers under the random Surge-2 demand condition using three evaluation metrics: average queue length, average waiting time, and average speed.

\begin{table}[h!]
\centering
\scriptsize
\setlength{\tabcolsep}{3pt}
\caption{As Table~\ref{tab:surge1_summary}, under Surge-2 (random per-flow +10\%--60\%).}
\label{tab:surge2_summary}
\resizebox{\columnwidth}{!}{%
\begin{tabular}{|l|c|c|c|c|c|}
\hline
\textbf{Controller} & \textbf{Avg Queue} & \textbf{Wait} & \textbf{Speed} & \textbf{$\Delta$ vs.\ Ref.} & \textbf{Hedges'} \\
 & \textbf{(Mean $\pm$ Std) $\downarrow$} & \textbf{(s) $\downarrow$} & \textbf{(m/s) $\uparrow$} & \textbf{(95\% CI)} & \textbf{$g$} \\
\hline
PPO (Queue AE-16D) & 29.69 $\pm$ 4.64 & 0.74 & 9.60 & [-2.33, +4.80] & +0.33 \\
PPO (Delay)        & 35.49 $\pm$ 4.20 & 0.81 & 9.55 & [+3.20, +10.86] & +2.05 \\
\textbf{PPO (Queue AE-8D)}$^{\dagger}$ & \textbf{28.46 $\pm$ 1.24} & 0.67 & \underline{9.69} & -- & -- \\
PPO (Pressure)     & 36.76 $\pm$ 4.18 & 0.80 & 9.60 & [+5.10, +11.19] & +2.44 \\
PPO (Full State)   & 31.24 $\pm$ 4.09 & 0.69 & 9.57 & [-0.12, +5.69] & +0.83 \\
PPO (Queue AE-19D) & \underline{28.96 $\pm$ 3.02} & 0.66 & 9.68 & [-2.32, +3.33] & +0.20 \\
PPO (Queue AE-32D) & 31.43 $\pm$ 3.19 & 0.80 & 9.53 & [+0.64, +5.31] & +1.11 \\
PPO (Queue)        & 32.67 $\pm$ 2.29 & 0.70 & 9.66 & [+2.02, +6.41] & +2.07 \\
PPO (Queue AE-4D)  & 29.27 $\pm$ 1.95 & 0.66 & 9.66 & [-0.90, +3.27] & +0.45 \\
RESCO-DQN          & 35.67 $\pm$ 7.97 & \textbf{0.37} & \textbf{9.84} & [+2.80, +13.61] & +1.14 \\
Actuated (SUMO)    & 36.76 $\pm$ 0.52 & 0.91 & 9.59 & [+7.21, +9.30] & +7.91 \\
Dynamic Webster    & 109.91 $\pm$ 0.81 & 6.70 & 7.87 & [+80.28, +82.58] & +70.27 \\
\hline
\end{tabular}%
}
\end{table}

\begin{figure}[htbp]
\centering
\includegraphics[width=0.48\textwidth]{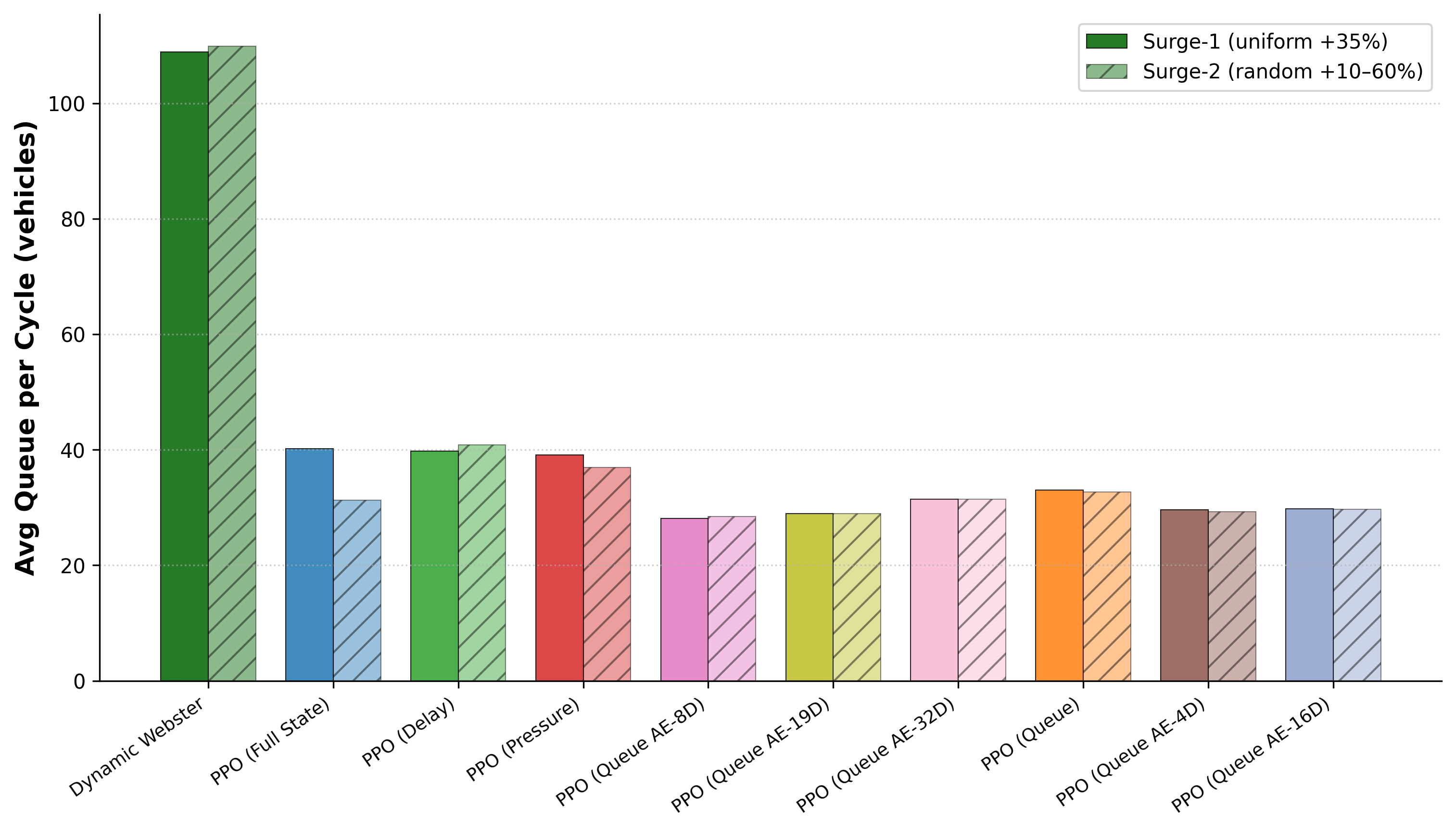}
\caption{Side-by-side comparison of average queue length per controller under Surge-1 (uniform +35\%, solid bars) and Surge-2 (random +10\%-60\%, hatched bars).}
\label{fig:surge_comparison}
\end{figure}

The ranking under Surge-2 closely mirrors Surge-1, and Webster remains far behind all RL variants (67\%--74\% lower mean queue for RL, computed from the entries of Table \ref{tab:surge2_summary}). \textcolor{black}{AE-8D has the lowest mean queue and the lowest standard deviation among the RL configurations in both surges, but its differences to PPO (Full State), AE-19D, AE-4D and AE-16D have intervals that include zero, so this is a descriptive observation only.} Average queue lengths under Surge-2 are close to those under Surge-1 for most configurations, since the mean per-flow increase in Surge-2 ($\approx$+35\%) matches Surge-1 (Fig.~\ref{fig:surge_comparison}).

\section{Conclusions and Future Scope} \label{sec:conclusion}
This paper presented an RL based adaptive TSC framework whose reward depends only on approach-level queue length, and applied it to a conventional, lane-based homogeneous traffic (LBHO) intersection and to a real, non-orthogonal junction with laneless heterogeneous (LLHE) traffic. At the LLHE site, all RL configurations had 66--73\% lower mean queue than our Dynamic Webster implementation under nominal flows and 63--74\% lower under two higher-demand scenarios.
At the LBHO site, the lowest-mean configuration  had about 23\% lower mean queue than Webster and the unselected Full State about 15\% lower, although most PPO variants were statistically indistinguishable at five seeds. Under identical state representations, we detected no difference between the queue based reward and delay- or pressure based rewards; the queue reward needs only approach-level queue counts, which motivates its use under laneless traffic, but robustness to estimated queues was not tested. We also release what is, to our knowledge, the first open laneless, heterogeneous-vehicle SUMO configuration integrated with RL signal control on a current SUMO version.

This study was scoped to a single, isolated intersection under each traffic regime to isolate the effects of state representation and reward design from inter-signal coordination. Future work will extend the framework to multi-agent RL for coordinated control of multiple signals along a corridor, and will evaluate performance under combined stress scenarios involving simultaneous demand surges and incident-driven capacity reductions.It should also add fixed-time and max-pressure baselines, test the queue reward under noisy or quantised queue estimates, compare autoencoder encodings with PCA, random projections and end-to-end encoders, include a leave-one-block-out ablation of the Full State, and repeat the main comparisons with more seeds.

\bibliographystyle{IEEEtran}
\bibliography{ref}

\appendices
\section{Operational Constraint Derivations}
\label{app:constraints}

This appendix gives the full derivation of the green-time and yellow-time bounds summarized in the Operational Constraints subsection of Section~\ref{sec:formulation}.

Assuming a pedestrian walking speed of $v_p = 1.2\,\text{m/s}$ \cite{manual2000highway}, the minimum green time, $g_{\text{min}}$, required for safe pedestrian crossing is approximately 10 seconds. The maximum green time, $g_{\text{max}}$, for each phase is limited to 40 seconds based on the codal provision [IRC SP: 41-1994] \cite{congress1994guidelines}. These values define the operational bounds for green time adjustments. Actions are subject to the following constraint: they are only available if the current green time $g_i$ for the active phase exceeds a minimum threshold, specifically 10 seconds. Until this threshold is reached, the controller is not allowed to intervene and no action is taken (i.e., the system continues the green phase by default). In addition, all green time adjustments must respect operational bounds:
\begin{equation}
g_{\min} \leq g_i \leq g_{\max},
\end{equation}
where $g_{i}$ is the green time for the phase $i$, and $g_{\min}$ and $g_{\max}$ denote the minimum and maximum permissible green times for a given phase.

The yellow time is calculated using the dilemma zone model, which seeks to prevent situations where a vehicle neither stops safely nor clears the intersection in time \cite{gartner2002traffic}. The minimum yellow time $Z_{\min}$ is determined using the equation
\begin{equation}
Z_{\min} = t_f + \frac{W + L_{\text{veh}}}{u_0} + \frac{u_0}{2a},
\end{equation}
where $t_f$ is the driver reaction time, $W$ is the intersection width, $L_{\text{veh}}$ is the effective vehicle length, $u_0$ is the approach speed of the vehicle, and $a$ is the comfortable deceleration rate. Assuming $t_f = 1\,\text{s}$, $W = 12.4\,\text{m}$, $L_{\text{veh}} = 10.2\,\text{m}$, $u_0 = 11.11\,\text{m/s}$, and $a = 3.53\,\text{m/s}^2$, the minimum yellow time is found to be approximately $Z_{\min} \approx 5\,\text{s}$.

\section{\textcolor{black}{Reward Weighting Sweep}}
\label{app:alpha}
\textcolor{black}{Table~\ref{tab:alpha_sweep} lists results for alternative weightings $(\alpha_{abs}, \alpha_{red})$ at the LLHE site, for reference. The $(0.4, 0.6)$ row is the setting used in the main experiments.}

\begin{table}[h!]
\centering
\scriptsize
\caption{Effect of reward weighting $(\alpha_{abs}, \alpha_{red})$ on average total queue length per cycle, LLHE site, raw-queue state, nominal flows. \textcolor{black}{Run with SUMO 1.22.0.} Bold marks the setting used in the main experiments.}
\label{tab:alpha_sweep}
\begin{tabular}{|c|c|c|c|}
\hline
\textbf{$(\alpha_{abs}, \alpha_{red})$} & \textbf{Avg.\ Queue/Cycle} & \textbf{Wait (s)} & \textbf{Speed (m/s)} \\
 & \textbf{($\mu \pm \sigma$) $\downarrow$} & \textbf{$\downarrow$} & \textbf{$\uparrow$} \\
\hline
(1.0, 0.0) & 32.45 $\pm$ 3.71 & 0.80 & 9.53 \\
(0.8, 0.2) & 34.10 $\pm$ 3.46 & 0.87 & 9.51 \\
(0.6, 0.4) & 32.41 $\pm$ 2.51 & 0.78 & 9.58 \\
\textbf{(0.4, 0.6)} & \textbf{30.09 $\pm$ 1.59} & \textbf{0.71} & \textbf{9.61} \\
(0.2, 0.8) & 35.12 $\pm$ 2.76 & 0.90 & 9.50 \\
(0.0, 1.0) & 32.55 $\pm$ 3.84 & 0.80 & 9.56 \\
\hline
\end{tabular}
\end{table}

\section{\textcolor{black}{Cumulative State Build-up}}
\label{app:buildup}
\textcolor{black}{This appendix reports a supplementary cumulative build-up of the Full State (Eq.~\eqref{eq:state_full}): each level adds one block to the previous level, and only the observation changes between levels. The levels are}
\begin{itemize}
\item \textcolor{black}{3-D: temporal features $T_c$, $t_p$, $N_{\text{cycles}}$;}
\item \textcolor{black}{7-D: $+$ phase one-hot $\mathbf{P}$;}
\item \textcolor{black}{11-D: $+$ per-approach queue $\mathbf{Q}$;}
\item \textcolor{black}{15-D: $+$ queue change $\Delta\mathbf{Q}$;}
\item \textcolor{black}{19-D: $+$ green times $\mathbf{g}$, i.e.\ the Full State.}
\end{itemize}
\textcolor{black}{All levels use the same PPO settings, reward and seeds as the corresponding Full-State runs. At the LLHE site the third and fourth entries of $\mathbf{P}$ and $\mathbf{g}$ (the intersection has two green phases) and the queue entries of one approach are constant, so 3, 5, 8, 11 and 13 features vary at the five levels. Because the design is cumulative addition rather than leave-one-block-out ablation, it cannot isolate the contribution of any single block. Under the queue reward at the LLHE site (Table~\ref{tab:dim_buildup_hetero}), only the step from 3-D to 7-D (adding the phase one-hot) has an interval that excludes zero (35.67 $\to$ 30.73, CI $[-8.16, -1.63]$); later steps have intervals that include zero. At the LBHO site, under the queue reward (Table~\ref{tab:dim_buildup}) no step has an interval excluding zero, and under the delay reward (Table~\ref{tab:dim_buildup_delay}) only the last step (adding $\mathbf{g}$, 58.97 $\to$ 70.31) does, in the direction of a higher queue. We draw no general conclusion from these results.}

\begin{table}[t]
\centering
\caption{Feature build-up within the 19-D Full State, homogeneous lane based site, queue reward (5 seeds). Each level adds one block of Eq.~\eqref{eq:state_full} to the previous level; $\Delta$ (bootstrap 95\% CI) and Hedges' $g$ compare each level with the level above it, paired by seed; $^{\dagger}$ marks the first level, which has no comparison.}
\label{tab:dim_buildup}
\resizebox{\columnwidth}{!}{%
\begin{tabular}{clccccc}
\toprule
Dim & Features added & Avg.\ queue (veh) & Wait (s) & Speed (m/s) & 95\% CI of $\Delta$ & $g$ \\
\midrule
3$^{\dagger}$ & temporal ($T_c$, $t_p$, $N_{\text{cycles}}$) & 54.58 $\pm$ 6.49 & 25.87 & 2.810 & -- & -- \\
7 & $+$ phase one-hot $\mathbf{P}$ & 60.60 $\pm$ 10.72 & 20.61 & 2.114 & [-6.62, +18.38] & +0.61 \\
11 & $+$ per-approach queue $\mathbf{Q}$ & 62.73 $\pm$ 8.30 & 26.86 & 2.277 & [-11.46, +15.55] & +0.20 \\
15 & $+$ queue change $\Delta\mathbf{Q}$ & 53.70 $\pm$ 16.98 & 20.57 & 2.760 & [-22.48, +0.44] & -0.61 \\
19 & $+$ green times $\mathbf{g}$ (= Full State) & 58.86 $\pm$ 5.68 & 26.46 & 2.378 & [-7.28, +17.60] & +0.37 \\
\bottomrule
\end{tabular}}
\end{table}

\begin{table}[t]
\centering
\caption{Feature build-up within the 19-D Full State, homogeneous lane based site, delay reward (5 seeds). Layout and statistics as in Table~\ref{tab:dim_buildup}}
\label{tab:dim_buildup_delay}
\resizebox{\columnwidth}{!}{%
\begin{tabular}{clccccc}
\toprule
Dim & Features added & Avg.\ queue (veh) & Wait (s) & Speed (m/s) & 95\% CI of $\Delta$ & $g$ \\
\midrule
3$^{\dagger}$ & temporal ($T_c$, $t_p$, $N_{\text{cycles}}$) & 59.09 $\pm$ 1.48 & 21.80 & 2.386 & -- & -- \\
7 & $+$ phase one-hot $\mathbf{P}$ & 63.90 $\pm$ 6.22 & 23.56 & 1.985 & [-0.44, +9.43] & +0.96 \\
11 & $+$ per-approach queue $\mathbf{Q}$ & 57.17 $\pm$ 8.06 & 26.01 & 2.749 & [-13.62, +0.29] & -0.84 \\
15 & $+$ queue change $\Delta\mathbf{Q}$ & 58.97 $\pm$ 9.97 & 24.46 & 2.365 & [-4.31, +9.34] & +0.18 \\
19 & $+$ green times $\mathbf{g}$ (= Full State) & 70.31 $\pm$ 1.85 & 29.58 & 1.672 & [+3.36, +19.15] & +1.43 \\
\bottomrule
\end{tabular}}
\end{table}

\begin{table}[t]
\centering
\caption{Feature build-up within the 19-D Full State, heterogeneous LLHE site, queue reward (5 seeds). Layout and statistics as in Table~\ref{tab:dim_buildup}.}
\label{tab:dim_buildup_hetero}
\resizebox{\columnwidth}{!}{%
\begin{tabular}{clccccc}
\toprule
Dim & Features added & Avg.\ queue (veh) & Wait (s) & Speed (m/s) & 95\% CI of $\Delta$ & $g$ \\
\midrule
3$^{\dagger}$ & temporal ($T_c$, $t_p$, $N_{\text{cycles}}$) & 35.67 $\pm$ 2.16 & 0.79 & 9.549 & -- & -- \\
7 & $+$ phase one-hot $\mathbf{P}$ & 30.73 $\pm$ 3.39 & 0.83 & 9.513 & [-8.16, -1.63] & -1.57 \\
11 & $+$ per-approach queue $\mathbf{Q}$ & 29.91 $\pm$ 3.20 & 0.77 & 9.538 & [-2.80, +1.16] & -0.22 \\
15 & $+$ queue change $\Delta\mathbf{Q}$ & 31.29 $\pm$ 3.68 & 0.79 & 9.526 & [-2.24, +5.57] & +0.36 \\
19 & $+$ green times $\mathbf{g}$ (= Full State) & 30.56 $\pm$ 2.37 & 0.69 & 9.617 & [-4.29, +2.83] & -0.21 \\
\bottomrule
\end{tabular}}
\end{table}

\section{Supplementary Figures}
\label{sec:appendix}

\end{document}